\documentclass[12pt,fleqn]{article}
\usepackage{amsmath,amssymb,amsthm,epsfig,graphics}

\renewcommand{\bar}{\overline} 
\newcommand{\ri}{{\mathrm i}}
\newcommand{\p}{\partial}
\newcommand{\bea}{\begin{array}}
\newcommand{\eea}{\end{array}}

\catcode`@=11 \long
\def\@caption#1[#2]#3{\par\addcontentsline{\csname
ext@#1\endcsname}{#1} {\protect\numberline{\csname
the#1\endcsname}{\ignorespaces #2}} \begingroup \small
\@parboxrestore \@makecaption{\csname fnum@#1\endcsname}
{\ignorespaces #3}\par \endgroup} \catcode`@=12

\renewcommand{\bar}{\overline}

\newcommand{\N}{\mathbb{N}}

\newcommand{\la}{\label}

\catcode`@=11 \long
\def\@caption#1[#2]#3{\par\addcontentsline{\csname
ext@#1\endcsname}{#1} {\protect\numberline{\csname
the#1\endcsname}{\ignorespaces #2}} \begingroup \small
\@parboxrestore \@makecaption{\csname fnum@#1\endcsname}
{\ignorespaces #3}\par \endgroup} \catcode`@=12

\begin{document}

\allowdisplaybreaks
 \begin{titlepage} \vskip 2cm

\begin{center} {\Large\bf Group classification of (1+3)-dimensional
Schr\"odinger equations with position dependent mass}
\footnote{E-mail: {\tt nikitin@imath.kiev.ua} } \vskip 3cm {\bf {A.
G. Nikitin } \vskip 5pt {\sl Institute of Mathematics, National
Academy of Sciences of Ukraine,\\ 3 Tereshchenkivs'ka Street,
Kyiv-4, Ukraine, 01004\\}}\end{center} \vskip .5cm \rm

\begin{abstract}
Kinematical invariance groups of the 3d Schr\"odinger equations with
position dependent masses (PDM) and arbitrary potentials are
classified. It is shown that there exist 94 classes of such
equations  defined up to the generic equivalence group, and 70
classes defined up to the equivalence groupoid. The maximally
extended kinematical invariance algebras of such equations appears
to be eight dimensional.

The specific symmetries connected with the presence of the ambiguity
parameters are discussed and  an extended class of systems which
keep their forms for arbitrary or particular changes of these
parameters is specified.

The exact solution of the selected PDM Schr\"odinger equation is
presented. This equation describes a deformed 3d isotropic harmonic
oscillator and possesses extended  continuous symmetries and  hidden
supersymmetries with two different superpotentials as well.
\end{abstract}
\end{titlepage}
\section{Introduction\label{int}}
Group classification of Schr\"odinger equations with arbitrary
potentials had been carried out in the seventieth of the previous
century in papers \cite{Hag}, \cite{Nied}, \cite{And} and
\cite{Boy}. This classification is a corner stone in the group theoretical grounds of quantum mechanics. It presents a
priori information about all admissible Lie symmetries of one
particle quantum mechanical systems, and is an important (and
necessary) step in investigation of various generalized symmetries
of Schr\"odinger equations.

In contrary, the group classification of Schr\"odinger equations
with position dependent mass (PDM) is still far from the
completeness. It can be considered as a challenge for experts in
group analysis of differential equations,  taking into account the
fundamental role played by such equations in  modern theoretical and
mathematical physics. There are many papers devoted to  PDM
Schr\"odinger equations with various particular symmetries, see,
e.g., \cite{11},  \cite{Cru}, \cite{Mul1}, \cite{Mul2}. But the
basic symmetries of these equations with respect to continuous
transformation groups are still unknown.

The first steps  in  the systematic study of symmetries of the PDM
Schr\"odinger equations was made apparently in papers \cite{NZ} and
\cite{NZ2}. Namely, in paper \cite{NZ}  the stationary 3$d$ PDM
Schr\"odinger equations with different symmetry groups have been
classified. In other words, all such equations admitting first order
integrals of motion were presented there.

In paper \cite{NZ2} we discuss symmetry properties of
non-stationary PDM Schr\"odinger equations with arbitrary number of
spatial variables and present the completed group classification of
such equations with two spatial variables. The number of the
equations with non-equivalent symmetry groups appears to be rather
restricted. More exactly, there exist only seven classes of such
equations, three of which are defined up to arbitrary functions and
four -- up to arbitrary parameters.

The 3d PDM systems admitting second order integrals of motion were
classified in paper \cite{N2}. The classification presented there is
completed but restricted to the PDM Schr\"odinger equations
invariant w.r.t. the rotation group.

The natural next step is to extend the analysis presented in
\cite{NZ} and \cite{NZ2} to the time dependent PDM Schr\"odinger
equations with three spatial variables, which is the most interested
from the physical point of view. Just such equations are discussed
in the present paper. We make the completed group classification of
them and specify 94 classes of equations with various symmetry
groups. In other words, we classify  all continuous symmetries which
can be admitted by PDM systems and constructively describe all
systems with non-trivial symmetries up to equivalence
transformations.

The specificity of PDM Schr\"odinger equations is that they include
the ambiguity parameters. We consider also the specific symmetries
connected with the presence of the ambiguity parameters and specify
an extended class of systems which keep their forms for arbitrary or
special changes of these parameters.

In contrast with the case of constant mass, PDM Schr\"odinger
equations have less extended symmetries.  However, there is a lot of
inequivalent equations admitting six, seven, or  eight dimensional
symmetry algebras. Such symmetry is sufficient to make equation
separable and even exactly solvable.

Quantum mechanical systems with extended Lie symmetries in many
cases are also supersymmetric. We do not study supersymmetric
aspects of PDM Schr\"odinger equations here, but consider an example
of a PDM system  which possesses  both the  extended Lie symmetry
and supersymmetry. This system  which we call deformed isotropic
oscillator  is shape invariant,  exactly solvable and has a more
general discrete energy spectrum (including an additional spectral
parameter) than the standard 3d isotropic oscillator. This system
together with many other supersymmetric and exactly solvable ones
are included   as a particular cases in our classification.

 Since the
Schr\"odinger equation with a constant mass appears as a particular
case in our analysis, we revised also the classical results
presented in the Boyer paper \cite{Boy} and find  few  systems which
were not presented in this classical work.

\section{Time dependent PDM Schr\"odinger equations}
We will discuss PDM Schr\"odinger equations of the following generic
form
\begin{gather}\la{seq}L\psi\equiv\left(\ri\frac{\p}{\p
t}-H\right)\psi=0\end{gather} where $H$ is the PDM Hamiltonian
\begin{gather}\la{A1}  H=\frac14(m^\alpha p_a m^{\beta}p_am^\gamma+
m^\gamma p_a m^{\beta}p_am^\alpha)+ \hat V.\end{gather} Here
$p_a=-\ri\frac{\p}{\p x_a}$, $m=m({\bf x})$ and $\hat V=\hat V({\bf
x})$ are the mass and potential depending on spatial variables ${\bf
x}=(x_1,x_2,x_3)$, and summation w.r.t. the repeating indices $a$ is
imposed over the values $a=1,2,3$.

The first term in the r.h.s. of equation (\ref{A1}) is interpreted as a kinetic energy term.
We use the standard representation
  with $\alpha, \beta$  and $\gamma$ being the
ambiguity parameters satisfying the condition $\alpha+\beta+\gamma=-1$ \cite{Roz}.

The choice of values of the ambiguity parameters can be motivated by
physical reasons, see a short discussion of this point in
\cite{NZ2}. Mathematically, we can treat two Hamiltonians (\ref{A1})
with different values of these parameters as similar provided
their potential terms  are not fixed. Indeed, Hamiltonian (\ref{A1})
can be rewritten in the form including another values of the
ambiguity parameters marked by tildes:
\begin{gather}\la{A12}  H=\frac14(m^{\tilde\alpha} p_a
m^{\tilde \beta}p_am^{\tilde\gamma}+m^{\tilde\gamma} p_a
m^{\tilde\beta}p_am^{\tilde\alpha})+  V\end{gather} provided $\hat
V$ and $V$ are connected by the following relation:
\begin{gather}\la{sim} V-\hat V=\frac14(\tilde\beta-\beta)f_{aa}+ (\alpha\gamma-\tilde\alpha\tilde\gamma)\frac{f_a f_a}{2f}\end{gather}
were $f=\frac1m,\ f_a=\frac{\p f}{\p x_a}$ and $f_{aa}=\Delta f=\frac{\p f_a}{\p x_a}$.
In particular, we can choose $\tilde\alpha=\tilde\gamma=0$ and reduce
Hamiltonian (\ref{A1}) to the following form:
\begin{gather}\la{A11}H= \frac12p_a fp_a+  V.\end{gather}

We will restrict ourselves to classification of equations
(\ref{seq}) with Hamiltonians in the form (\ref{A11}). The
classification results can be easily reformulated for systems with
generic Hamiltonian (\ref{A1}) using relation  (\ref{sim}) with
$\tilde\alpha=\tilde\gamma=0,\ \tilde\beta=-1$. In other words,
potentials of equations in generic form (\ref{seq}) can be expressed
via potentials $ V$ by the following formula:

\begin{gather}\la{simsim}\hat V=V-\frac{\alpha+\gamma}4f_{aa}-\alpha\gamma
\frac{f_a f_a}{2f}.\end{gather}

An interesting special cases of relation (\ref{sim}) appears if
$\hat V=V$
 or $\hat V=V+Const.$ If so, the potentials in  Hamiltonians
 (\ref{A1}) and (\ref{A12}) coincide, exactly in the first case and up a
 constant term in  the second. Since Hamiltonians (\ref{A1}) and (\ref{A12})
  are equal one to another, their kinematical parts are equal as well.
  Then the related  equation (\ref{sim}) represents equivalence relations
  for the ambiguity parameters, see Section 8 for discussion of
  this point.

\section{Determining equations for symmetries }
Let us search for symmetries of equations (\ref{seq}), (\ref{A11})
with respect to Lie groups of transformations of dependent and
independent variables. Since these equations are linear, they will
not be changed if we add to $\psi$  arbitrary particular solutions
$\psi_p$. The changes
\begin{gather}\la{ququ}\psi\to\psi+\psi_p\end{gather}
keep the equations
invariant and form the infinite group of their symmetries.

Our problem is to find all symmetries additional to (\ref{ququ}). It
can be reduced to searching for the first order differential
operators:
\begin{gather}\label{so}
    Q=\xi^0\partial _t+\xi^a\partial_a+\tilde\eta\equiv \xi^0\partial _t+
    \frac12\left(\xi^a\partial_a+\partial_a\xi^a\right)+\ri\eta,
\end{gather}
associated with  generators of the continuous symmetry groups of
equation (\ref{seq}), (\ref{A11}). In (\ref{so}) $\tilde
\eta=\frac12\xi^a_a+\ri\eta,\ \ $ $\xi^0$, $\xi^a$ and $\eta$ are
functions of independent variables, whose explicit form can be found
from the following operator equation
\begin{gather}\la{ic}QL-LQ=a L\end{gather}
where $a$ is one more unknown function of $t$ and $\bf x$.

The condition  (\ref{ic}) generates a system of differential
equations for functions  $\xi^0, \xi^a, \eta, f, \hat V $ and $a$:
\begin{gather}
\dot\xi^0=-a, \quad  \xi^0_a=0,\la{de1}\\
\la{de5} \xi^b_{a}+\xi^a_{b}-\frac{2}3\delta_{ab}\xi^i_i=0,
\\\label{de6} \xi^if_i-a f=\frac23 f\xi^i_i,\\ \dot\xi^a+\eta_a
f=0,\label{de7}\\
\label{de8}\xi^aV_a+\frac14\xi^b_{ba}f_a=a V+\dot\eta.\end{gather}
Here $\delta^{ab}$ is the Kronecker symbol, the dot denotes the
derivation w.r.t. the time variable whilst derivations w.r.t. the
spatial variables are denoted by subindices: $\dot \xi^0=\frac{\p
\xi^0}{\p t}, \ f_{a}=\frac{\p f}{\p x_a}$, etc.

 We will not deduce the system of the determining equations (\ref{de5})--(\ref{de8}) here since it is nothing but a special case of the system corresponding to the PDM Schr\"{o}dinger equation with {\it arbitrary} number of spatial variables presented in \cite{NZ2}.

Equation (\ref{de5}) defines the conformal Killing vector
whose generic form  is given by the following formula:
\begin{gather}\la{kil} \xi^a=r^2\lambda_a-2x_i\lambda_ix_a
+\omega x_a+\varepsilon_{abc} x_b\theta_c+\nu_a\end{gather} where
$r^2=x_1^2+x_2^2+x_3^2$,  $\varepsilon_{abc}$ is the totally
antisimmetric unit tensor while $\lambda_a,\ \omega,\ \theta_c$ and
$\nu_a$ are parameters which do not depend on $x$ but in general can
depend on $t$. Notice that, in accordance with (\ref{de1}), function
$\xi^0$ does not depend on ${\bf x}=(x_1,x_2,x_3)$.

Considering the remaining determining equations (\ref{de6})--(\ref{de8}) and using  (\ref{kil}), we reduce them to the following form:
\begin{gather}\la{5.1}\dot \xi^a+\eta_af=0,\\
\la{5.2}\xi^a\hat f_a=a+2\omega-4\lambda_bx_b,\\
\la{5.3}\xi^aV_a-\frac32\lambda_af_a=a V+\dot \eta\end{gather} where
the summation is imposed over the repeating indices $a$ over the
values $a=1, 2, 3$.

Thus the group classification of equations (\ref{seq}) is reduced to finding
inequivalent solutions of equations (\ref{5.1})--(\ref{5.3}) where $\xi^a$ are given by formula (\ref{kil}).

\section{Algebraic content of symmetry operators and \\ equivalence relations}
 Let us discuss  the algebraic structure of symmetry operators  (\ref{so}), (\ref{kil}). It happens these operators can be represented as:
\begin{gather}\label{Q1} Q=\xi^0\p_0+\lambda_i K_{i}+\theta_iL_i+\omega D+\nu_iP_i+\eta\end{gather}
where
\begin{gather}\label{QQ}\begin{split}&
 P_{i}=p_{i}=-i\frac{\partial}{\partial x_{i}},\quad L_{i}=\varepsilon_{ijk}x_jp_k, \\&
D=x_n p_n-\frac{3\ri}2,\quad K_{i}=x_nx_n p_i -2x_iD.\end{split}
\end{gather}

The evident solutions of relations (\ref{de1})--(\ref{de8}) valid for {\it arbitrary} $f$ and $V$ are $\xi^a=\eta=0, $ $\xi^0=1$. They correspond to the following symmetry operator $Q=P_0$:
\begin{gather}\la{P0}P_0=\ri\p_t\end{gather}
which generates shifts of the time variable.

 Operators (\ref{QQ}) form a basis of the Lie algebra c(3) of conformal group  defined in the 3$d$ Euclidean space. This algebra is isomorphic to so(1,4), i.e., to the Lie algebra of the Lorentz group in 1+4 dimensional space. This isomorphism can be fixed by choosing the new basis of algebra c(3):
\begin{gather}M_{ab}=\varepsilon_{abc}L_c,\ M_{0a}=\frac12(K_a+P_a),
\ M_{4a}=\frac12(K_a-P_a),\  M_{04}=D\label{iso}\end{gather}
where $M_{\mu\nu}=-M_{\nu\mu}$ with $\mu, \nu=0,1,2,3,4$ are basis
elements of algebra so(1,4), satisfying the following commutation
relations:
\begin{gather}\label{13}[M_{\mu\nu},M_{\lambda\sigma}]=\ri (g_{\mu\sigma}M_{\nu\lambda}+g_{\nu\lambda}M_{\mu\sigma}-
g_{\mu\lambda}M_{\nu\sigma}-g_{\nu\sigma}M_{\mu\lambda})\end{gather}
were $g_{\mu\nu}=diag(1,-1,-1,-1,-1)$.

Let us also note that  the conformal group C(3) is the equivalence group of Hamiltonian
(\ref{A11}) \cite{NZ}. Indeed, just conditions (\ref{de5}) 
are necessary and sufficient for keeping the general structure of Hamiltonian (\ref{A11}) while functions $f$ and $V$ can be changed.  In other words, transformations of spatial variables belonging to group C(3)
 do not change
the generic form of the Hamiltonian and can be used to simplify it and  symmetries (\ref{so}) as well.

To
present a way of such simplification we expand symmetry operators
via basis elements of algebra so(1,4):
\begin{gather}\la{013}Q=\xi^0\p_0+\tau_{\mu\nu}M_{\mu\nu}+\eta\end{gather} where
$\tau_{\mu\nu}$ are functions of time and the summation over the
repeating indices is imposed over the values 0, 1, 2, 3, 4.

We will see that all functions $\tau_{\mu\nu}$ should be
proportional to some fixed function, and so it is possible to reduce
the number of nonzero components of tensor $M_{\mu\nu}$ using its
transformation properties with respect to the group SO(1,4). For
example, let $\tau_{0a}=0$ for all $a=1,2,3,4,$ then we can restrict
ourselves to symmetries (\ref{013}) with the only nonzero
coefficient $\tau_{12}$ and then obtain the generic case with
nontrivial $\tau_{13}, \tau_{14}, \tau_{23}, \tau_{24}$ and
$\tau_{34}$, using a transformation belonging to SO(4)$\subset$ SO(1,4).

The completed description of all nonequivalent linear combinations
of tensors $M_{\mu\nu}$ satisfying relations (\ref{13}) was
presented in paper \cite{Pate}. We use these results to decouple the
 classification problem to the set of 
relatively simple subproblems, like it was done in \cite{NZ}.

We will search for potentials $V$ up to constant terms $C$, i.e.,
potentials $V$ and $V'=V+C$ will be treated as equivalent. To reduce
$V'$ to $V$ it is sufficient to make the transformation $\psi\to
e^{-\ri Ct}\psi$ in equation (\ref{A11}). In addition, rescaling the
time variable we can reduce to the unity any constant multiplier for
$f$.

Transformations discussed in the above belong to the generic equivalence
group of equations (\ref{A11}) and can be applied for any arbitrary
elements $f$ and $V$. However, for some particular masses and potentials
there are additional equivalence transformations which are listed in the
following formulae:
\begin{gather}\la{et02}\begin{split}&x_1\to\tilde
x_1=x_1\cos\left(\frac\mu{\lambda^2}\right)-x_2\sin\left(\frac\mu{\lambda^2}
\right),\\& x_2\to\tilde
x_2=x_2\cos\left(\frac\mu{\lambda^2}\right)+x_1\sin\left(\frac\mu{\lambda^2}
\right),\end{split}\end{gather}
\begin{gather}\la{et03}\begin{split}&x_1\to\tilde
x_1=x_1\cos\left(\ln(\omega)\right)-x_2\sin\left(\ln(\omega)
\right),\\& x_2\to\tilde
x_2=x_2\cos\left(\ln(\omega)\right)+x_1\sin\left(\ln(\omega)
\right),\quad t\to\tilde t={ t}\omega^{-1},\end{split}\end{gather}
\begin{gather}\la{et04}{\bf x}\to\tilde{\bf x}=
{\text e}^{\frac\mu{\lambda^2}}{\bf x},\end{gather}
\begin{gather}\la{et05}{\bf x}\to\tilde{\bf x}=
\omega^{\frac{-1}\sigma}{\bf x},\quad t\to\tilde
t={t}\omega^{-1},\end{gather}
\begin{gather}\la{et06}x_3\to\tilde x_3=
x_3-\omega,\quad t\to\tilde t={t}\omega^{-1},\end{gather}
\begin{gather}\la{et6}\begin{split}&x_1\to\tilde x_1={\sigma }^{-1},
\quad x_2\to\tilde x_2={\sigma}^{-1},\quad x_3\to\tilde x_3=
{\sigma}^{-1}( x_3+2\ln(\sigma))\end{split}\end{gather}
\begin{gather}\la{et2}\begin{split}&x_1\to
\tilde x_1=x_1\cos\left(\frac{\mu t^2}2\right)+
x_2\sin\left(\frac{\mu t^2}2\right),\\& x_2\to\tilde x_2=
x_2\cos\left(\frac{\mu t^2}2\right)-x_1\sin\left(\frac{\mu t^2}2\right),
\\&
\psi(t,x_1,x_2,x_3)\to\tilde\psi(t,\tilde x_1,\tilde
x_2,x_3)=\text{e}^{\ri\mu t(\frac{\mu
t^2}3-\Theta)}\psi(t,x_1,x_2,x_3),\\&
 V\to V+\mu\tilde\Theta,\quad \tilde\Theta=
 \arctan\left(\frac{\tilde x_2}{\tilde x_1}\right),\quad
 \Theta=\arctan\left(\frac{x_2}{x_1}\right),\end{split}\end{gather}
 \begin{gather}\la{et5}\begin{split}&{\bf x}\to \tilde {\bf x}=
 {\bf x}\text{e}^{-\frac{\mu t^2}2},\quad
\psi(t,{\bf x})\to\tilde\psi(t,\tilde {\bf x})=\text{e}^{\frac{3\mu
t^2}2+\ri\mu t(\frac{\mu t^2}3- \ln(r))}\psi(t,{\bf x}),\\&
 V\to V+\mu \ln(r),\end{split}\end{gather}
 \begin{gather}\la{et07}\begin{split}&{ x_3}\to \tilde { x_3}=
 { x_3}\text{e}^{-\frac{\mu t^2}2},\\&
\psi(t,{x_1,x_2, x_3})\to\tilde\psi(t, {x_1,x_2,\tilde x_3
})=\text{e}^{\frac{3\mu t^2}2+\ri\frac{\mu^2 t^3}3-\ri\mu t
\ln(x_3)}\psi(t,{x_1,x_2,x_3}),\\&
 V\to V+\mu \ln(x_3),\end{split}\end{gather}
\begin{gather}\la{et7}\begin{split}&x_3\to \tilde x_3=x_3+\ln(1+ t^2)+\ln\left(\frac{\mu}2\right),\quad t\to\tilde t=\frac2\mu\arctan( t),\\& \psi(t,x_1,x_2,x_3)\to\tilde\psi(\tilde t,x_1,x_2,\tilde x_3)=\text{e}^{\frac{-\ri\mu t}{(1+ t^2)}\text{e}^{- x_3}}\psi(t,x_1,x_2,x_3),\\&  V\to V+\frac{\mu^2}2\text{e}^{- \tilde x_3},\end{split}\end{gather}
\begin{gather}\la{et8}\begin{split}&x_3\to \tilde x_3=x_3+\ln(1- t^2)+\ln\left(\frac{\mu}2\right),\quad t\to\tilde t=\frac2\mu\text{arctanh}( t),\\& \psi(t,x_1,x_2,x_3)\to\tilde\psi(\tilde t,x_1,x_2,\tilde x_3)=\text{e}^{\frac{\ri\mu t}{(1- t^2)}\text{e}^{- x_3}}\psi(t,x_1,x_2,x_3),\\&  V\to V-\frac{\mu^2}2\text{e}^{- \tilde x_3},\end{split}\end{gather}
\begin{gather}\la{et9}\begin{split}&x_1\to \tilde x_1=x_1\cos(\Phi)-x_2\sin(\Phi),\quad \Phi=\frac{1}{\sigma}\ln\left(\frac{\omega\sigma(1+ t^2)}2\right),\\& x_2\to \tilde x_2=x_2\cos(\Phi)+x_1\sin(\Phi),\quad t\to\tilde t=\frac2{\omega\sigma}{\arctan(t)},\\& \psi(t,x_1,x_2,x_3)\to\tilde\psi(\tilde t,\tilde x_1,
\tilde x_2, x_3)=\text{e}^{\frac{-\ri\omega t}{\sigma(1+t)^2}
\text{e}^{-\sigma\Theta}}\psi(t,x_1,x_2,x_3),\\
&  V\to
V+\frac{\omega^2}2\text{e}^{-
\sigma\tilde\Theta},\end{split}\end{gather}
\begin{gather}\la{et10}\begin{split}&x_1\to \tilde x_1=x_1\cosh(\hat\Phi)-x_2\sinh(\hat\Phi),\quad \hat\Phi=\frac{1}{\sigma}\ln\left(\frac{\omega\sigma(1- t^2)}2\right),\\& x_2\to \tilde x_2=x_2\cosh(\hat\Phi)+x_1\sinh(\hat\Phi),\quad t\to\tilde t=\frac2{\omega\sigma}{\text{arctanh}(t)},\\& \psi(t,x_1,x_2,x_3)\to\tilde\psi(\tilde t,\tilde x_1,
\tilde x_2, x_3)=\text{e}^{\frac{\ri\omega t}{\sigma(1- t^2)}
\text{e}^{-\sigma\Theta}}\psi(t,x_1,x_2,x_3),\\
&  V\to
V-\frac{\omega^2}2\text{e}^{-
\sigma\tilde\Theta},\end{split}\end{gather}
\begin{gather}\la{et11}\begin{split}&{\bf x}\to \tilde {\bf x}=
{\bf x}(1+ t^2)^{\frac{1}{2\sigma}}, \quad t\to\tilde
t=\frac1{\omega\sigma}{\arctan(
t)},\\&\psi(t,{\bf x})\to\tilde\psi(\tilde t,\tilde{\bf x})=(1+ t^2)^{\frac{3}4}\text{e}^{\frac{-\ri\omega t}{2\sigma(1+
t^2)}({\bf x}^{2})^{-\sigma}}\psi(t,{\bf x}),\\&V\to
V+\frac{\omega^2}2(\tilde{\bf x}^{2})^{-\sigma},\end{split}\end{gather}

\begin{gather}\la{et12}\begin{split}&{\bf x}\to \tilde {\bf x}=
{\bf x}(1-(\sigma t)^2)^{\frac{1}{2\sigma}}, \quad t\to\tilde
t=\frac1{\sigma\omega}\text{arctanh}(\sigma t),
\\&\psi(t\mathrm{},{\bf x})\to\tilde\psi(\tilde t,\tilde{\bf x})=(1- t^2)^{\frac{3}4}\text{e}^{\frac{\ri\omega t}{2\sigma(1- t^2)}({\bf x}^{2})^{-\sigma}}
 \psi(t,{\bf x}),\\
 &V\to V-\frac{\omega^2}2 (\tilde{\bf x}^{2})^{-\sigma}.\end{split}\end{gather}

Transformations (\ref{et02})--(\ref{et12}) do not belong to the
equivalence group C(3) which is valid for arbitrary mass and
potential. However, they keep the generic form of equations
(\ref{seq}), (\ref{A11}) for some particular  $f$ and $V$. To find
these transformations we test  the pairwise non-equivalence of all
obtained equations, especially of those ones which have equivalent
invariance algebras. As a result  we discover the additional
equivalence transformations (\ref{et02})--(\ref{et12}) which extend
the equivalence group of equations (\ref{seq}) to the equivalence
gruppoid. 

Let us stress that there are  equivalence transformations additional to (\ref{et02})--(\ref{et07}) which can change arbitrary functions in mass and potential terms to other arbitrary functions. We do not enumerate them here since they are not essential.

\section{Dependence on time}
We already know the generic dependence of functions $\xi^0$ and
$\xi^a$ on $\bf x$ which is described by equation (\ref{Q1}). More sophisticated speculations are requested to
define their dependence on $t$. We will specify the following qualitatively different versions of symmetries.

1. Functions $\xi^a$ and $\xi^0$ are time dependent.  In this case we have the generic determining equations
(\ref{5.1})--(\ref{5.3}) with $a$ being a  function of $t$, whilst
$\eta$  in general depends on $t$ and $\bf x$. Such symmetries will be called symmetries of class 1.

2. Vector $\xi^a$ and function $\eta$ are time independent but
$\xi^0$ depends on time.  In this case function $a$ in equations
(\ref{5.2}) and (\ref{5.3}) is non-trivial. We will refer to the
corresponding symmetries as symmetries of class 2.

3. For symmetries of class 3 vectors $\xi^a$ are time independent
but $\xi^0$ and $\eta$ can depend on time. In accordance with (\ref{de1}) in
this case function $a$ is nontrivial if $\xi^0_t\neq0$.

4. Symmetries of class 4 -- vectors $\xi^a$ and their counterparts
$\xi^0$ and $\eta$ are time independent. In this case the generic
expression (\ref{kil}) for $\xi^a$ includes arbitrary constant
parameters $\lambda_a,\ \nu_a,\ \theta_a$ and $\mu$. In addition,
functions $\eta$ and $\xi^0$ in (\ref{Q1}) are reduced to constants,
and function $a$ in determining equations (\ref{5.2}) and
(\ref{5.3}) is zero.

 The fourth class in fact was considered in detail in paper \cite{NZ},
 were  the symmetries of the stationary Schr\"odinger equations with
 position  dependent mass have been classified. The only new feature in
 comparison with \cite{NZ}  is the additional symmetry with respect to shifts of time variables
  whose generator is given by equation (\ref{P0}).

 The second and the third cases are a bit more complicated thanks to the
 presence of new
 arbitrary functions of $t$, i.e.,  $a$ and $\eta$, in the determining
 equations. However, it can be handled in analogy with the fourth case
 considered in \cite{NZ}. To do it we can again exploit the results of
 Patera and Winternitz \cite{Pate} concerning the subgroup structure of
 group P(1,4), which is isomorphic to the equivalence group of equation
 (\ref{seq}). Any such subgroup can be confronted by a reduced version of
 the determining equations with a small number of arbitrary parameters
 $\lambda_a,\ \nu_a,\ \theta_a$ and $\omega$. Such equations with
 $a=0$ and $\dot\eta=0$ have been integrated in \cite{NZ} and can be
 solved also for $a$ nonzero.

 The first case looks as much more complicated. However, equation (\ref{5.1}) with
 non-trivial $\dot\xi^a$ appears to be a rather strong condition which
  essentially  reduces the number of the corresponding
 symmetry operators.

Let us start with the analysis of the first class of symmetries. To
evaluate the additional restrictions for $f$ generated by equation
(\ref{5.1}) with
 non-trivial $\dot\xi^a$ we
differentiate (\ref{5.2}) with respect to $t$ and obtain:
\begin{gather}\la{5.4}\dot\xi^a\hat f_a=\dot a+2\dot \omega-4\dot
\lambda_ax_a\end{gather} where $\hat f=\ln(f),\ \ \hat f_a=\frac{f_a}f$.

One more rather evident differential consequence of equations
(\ref{5.1}), (\ref{5.2}) and (\ref{kil}) is:
\begin{gather}\dot\xi^a\hat f_b-\dot\xi^b\hat f_a=4(\dot\lambda_ax_b-
\dot\lambda_bx_a)- 2\varepsilon_{abc}\dot\theta_c
.\la{5.6}\end{gather}

Thus if symmetries are time dependent we have the additional  conditions
for $f$ given above. Formulae (\ref{5.4}) and (\ref{5.6}) present a system
of four algebraic equations for three unknowns $\hat f_a$. Its compatibility
 condition reads:
\begin{gather}\la{5.8}\dot\xi^a\dot\theta_a=
4\varepsilon_{abc}\dot\xi^a\dot\lambda_bx_c.\end{gather}

Equation (\ref{5.8}) generates rather strong restrictions on
coefficient functions $\lambda_a,\ \nu_a,\ \theta_a$ and $\omega$.
Up to equivalence it admits   only three nontrivial solutions for
functions $\xi^a$ (see Appendix):
\begin{gather}\la{e71}\xi^1=-\dot\Phi x_2,\quad
\xi^2=\dot\Phi x_1,\quad  \xi^3=0, \\\xi^a= \omega x_a,
\quad \omega=\dot\Phi,\quad a=1,2,3, \la{e72}\\
\xi^1=\xi^2=0,\quad \xi^3=\dot\Phi\la{e73}\end{gather} where
$\Phi=\Phi(t)$ is a function of time. The corresponding functions
$\xi^0,$ and $a$ are:
\begin{gather}\la{e74}\xi^0=\Phi, \quad
a=-\sigma\dot\Phi\end{gather} where $\sigma$ is a constant.

Substituting (\ref{e71})--(\ref{e74}) into (\ref{5.1}) and
(\ref{5.3}) we come to the following consistency conditions of the
latter equations:
\begin{gather}\la{e75} \dddot \Phi=\kappa\dot\Phi, \quad \eta=F({\bf x})\dot\Phi.\end{gather}

The generic solutions of the first of equations (\ref{e75}) for $\dot\Phi$ can be
represented in the following form:
\begin{gather}\la{q7}\dot\Phi=at +b,\quad ab=0,\phantom{aaaaaa}
\texttt{if}\quad
\kappa=0,\\\la{q8}\dot\Phi=a\cos(\lambda t)+b\sin(\lambda t) \qquad
\texttt{if}\quad
\kappa=-\lambda^2<0,\\\la{q9}\dot\Phi=a\cosh(\lambda
t)+b\sinh(\lambda t) \quad \texttt{if}\quad
\kappa=\lambda^2>0\end{gather}
 where $a, b$ and $\lambda$ are arbitrary constants  and the condition $ab=0$ reflects the fact that for $a$
nonzero we can reduce $b$ to zero by an appropriate shift of the
time variable.

We specify the possible time dependence of functions $\xi^0$,
$\xi^a$, $a$ and $\eta$ for symmetries of the first type. For
symmetries of the second type we should deal with functions $\xi^a$
of generic form (\ref{kil}) with constant parameters $\lambda_a, \
\theta_a, \ \nu_a$ and $\omega$, while $\xi^0$ is a function of $t$.
The corresponding relations (\ref{de1}) and (\ref{5.2}) are
consistent iff this function is linear, i.e.,
\begin{gather}\xi^0=\sigma t, \quad a=-\sigma\la{f1}, \quad \dot\xi^a=0 \end{gather}
where $\sigma$ is a constant. In accordance with (\ref{5.1})
 function $\eta$ is a constant also.

For symmetries of the third type we have time independent $\xi^a$
discussed in the previous paragraph, and time dependent $\eta$
which, however, does not depend on $\bf x$. To satisfy equation
(\ref{5.3}) function $\eta$ should be linear in $t$, i.e., $
\eta=\mu t$ where $\mu$ is a constant.

Finally, for symmetries of the fourth type all functions satisfying
the determining equations (\ref{de1}) and (\ref{5.1})--(\ref{5.3})
by definition are time independent.

Thus we specify the possible dependence of coefficient functions
$\xi^0$, $\xi^a$ and $a$ on time and spatial variables and are in a
position to start the procedure of direct solving the determining
equations (\ref{5.1})--(\ref{5.3}). The  results of this solving are
presented in the following section while calculational detail can be
found in the Appendix.

\section{Classification results}
Let us present the results of the group classification of equations
(\ref{seq}). The number of such equations with essentially different
symmetries is rather extended.
The related position dependent masses are either fixed or arbitrary functions of reduced number of variables including parameters.

 The admissible fixed mass functions are presented in the following
 formulae:
\begin{gather}\la{m1}\begin{split}&
m=\tilde r^{-2}=(x_1^2+x_2^2)^{-1},\\& m=r^{-2},\quad
m=e^{-x_3},\quad m=(r^2\pm1)^{-2}\\&m=\tilde r^{-3},\ m=x_3^{-2}, \
m=x_3^{-3}.
\end{split}\end{gather} Just such functions correspond to the most
extended symmetries of equation (\ref{A11}).

Mass functions defined up to arbitrary  parameters also correspond to
highly symmetric PDM Schr\"odinger equations. They can have the following
forms:
\begin{gather}\la{m4}\begin{split}&m=r^{-2(1+\sigma)},\quad
m=\tilde r^{-2}e^{-\sigma \Theta},\quad m=\tilde
r^{-2-\sigma}e^{-\nu \Theta}\end{split}\end{gather}
 where  $\Theta=\arctan(\frac{x_2}{x_1}),$ $\sigma $ and $\nu$ are
 arbitrary nonzero parameters.  Masses with $m=r^{-2+\sigma}$ and
$m=r^{-2-\sigma}$ are equivalent.

In addition, in Table 6 we specify 24 mass terms defined up to arbitrary
functions.

\begin{center}Table 1\\
Systems with  power inverse PDM function $f=\tilde r^2=x_1^2+x_2^2.$
Parameter $\lambda$ takes  nonzero values, $\mu$ can be reduced to
zero applying transformation inverse to (\ref{et2}) for Items 1--6
and transformation (\ref{et02}) for Items 7--18. The invariance
algebras include the presented symmetries together with generator
(\ref{P0}) and the unit operator.

\end{center}
\begin{tabular}{r l l c }
\hline No&Potential $V$&Symmetries&$\begin{array}{c}\text{Invariance}\\\text{algebras}\end{array}$
\\
\hline
\\
1&$\begin{array}{l}G(\tilde r,x_3)+\mu\Theta,\\\texttt{where }
\Theta=\arctan(\frac{x_2}{x_1})\end{array} $ \vspace{2mm}
&$\begin{array}{l}A^1_1= t\left(L_3+\frac{\mu
t}2\right)-\Theta,\\
A^1_2=L_3+\mu t
\end{array}$& $\textsf{n}_{4,1}$\\
 2&$G(\tilde r)+\mu\Theta+\kappa
x_3$
\vspace{2mm}
&$\begin{array}{l}A^1_1,\ A^1_2, \ P_3+\kappa t
\end{array}$&$\textsf{n}_{5,4}$\\
3&$G(\frac{x_3}{\tilde r})+\mu\Theta+\nu\ln(\tilde r)$ \vspace{2mm}
&$\begin{array}{l}A^1_1,\ A^1_2, \
 D+\nu t
\end{array}$&$\textsf{n}_{5,4}$\\
4&$\begin{array}{l}G(\frac{r^2+1}{\tilde
r})+\mu\Theta+\lambda\Phi\\\text{where }
 \Phi=\arctan(\frac{r^2-1}{2x_3}) \end{array}$
 \vspace{2mm}
 &$\begin{array}{l}A^1_1,\ A^1_2, \
K_3-P_3-2\lambda t
\end{array}$&$\textsf{n}_{5,4}$\\
5&$\mu\Theta+\nu \ln(\tilde r)$ \vspace{2mm}
&$\begin{array}{l}A^1_1, \
 A^1_2,\ D+\nu t, \ P_3
\end{array}$&$\textsf{s}_{6,91}$\\
6&$\mu\Theta$ \vspace{2mm} &$\begin{array}{l}A^1_1, \
 A^1_2,\ P_3,\ D,\ \ K_3
\end{array}$&$\textsf{n}{_{4,1}}\oplus$ sl(2,R)
\\
7&$G(\tilde r,x_3)+\frac{\lambda^2}2\Theta^2+\mu\Theta$
\vspace{2mm} &$\begin{array}{l}B^1_1=\lambda\sin (\lambda t) L_3\\-
(\lambda^2\Theta+\mu)\cos (\lambda t),\vspace{2mm} \\
B^1_2=\lambda\cos (\lambda
t) L_3\\
+(\lambda^2\Theta+\mu)\sin (\lambda t)\end{array}$& $\textsf{s}_{4,7}$\\
8& \vspace{2mm} $G(\tilde
r)+\frac{\lambda^2}2\Theta^2+\mu\Theta+\kappa x_3$&$\begin{array}{l}B^1_1,\ B^1_2,\ P_3+\kappa t\end{array}$ &$\textsf{s}_{5,16}$\\
9& \vspace{2mm} $G(\frac{x_3}{\tilde
r})+\frac{\lambda^2}2\Theta^2+\mu\Theta +\nu\ln(\tilde
r)$&$\begin{array}{l}B^1_1,\ B^1_2,\
D+\nu t\end{array}$& $\textsf{s}_{5,16}$\\
10& \vspace{2mm} $G(\frac{r^2+1}{\tilde
r})+\frac{\lambda^2}2\Theta^2+\mu\Theta
+\lambda\Phi$&$\begin{array}{l}B^1_1,\ B^1_2,\
K_3-P_3-2\lambda t\end{array}$ &$\textsf{s}_{5,16}$\\
11& \vspace{2mm} $\frac{\lambda^2}2\Theta^2+\mu\Theta+\nu
\ln(\tilde
r)$& $\begin{array}{l}B^1_1,\ B^1_2,\ D+\nu t, \ P_3  \\
\end{array}$&$\textsf{s}_{6,230}$\\
12& \vspace{2mm} $\frac{\lambda^2}2\Theta^2+\mu\Theta$& $\begin{array}{l}B^1_1,\ B^1_2,\ P_3,\ D,\ K_3 \\
\end{array}$&$\textsf{s}_{4,7}\oplus$ sl(2,R) \\

13& \vspace{2mm} $G(\tilde
r,x_3)-\mu\Theta-\frac{\lambda^2}2\Theta^2$&$\begin{array}{l}C^1_1=\lambda\sinh
(\lambda t) L_3 \\+(\mu+\lambda^2\Theta)\cosh (\lambda t),\vspace{2mm} \\
C^1_2= \lambda\cosh (\lambda t) L_3\\+(\mu+\lambda^2\Theta)\sinh
(\lambda t)\end{array}$&$\textsf{s}_{4,6}$ \\
14& \vspace{2mm} $G(\tilde
r)-\mu\Theta-\frac{\lambda^2}2\Theta^2+\kappa
x_3$&$\begin{array}{l}C^1_1, \ C^1_2,\ P_3+\kappa t
\end{array}$&$\textsf{s}_{5,15}$
\\
15& \vspace{2mm} $G(\frac{x_3}{\tilde
r})-\mu\Theta-\frac{\lambda^2}2\Theta^2 +\nu\ln(\tilde
r)$&$\begin{array}{l}C^1_1, \ C^1_2, \ D+\nu t\end{array}$&
$\textsf{s}_{5,15}$\\
16& \vspace{2mm} $G(\frac{r^2+1}{\tilde
r})-\mu\Theta-\frac{\lambda^2}2\Theta^2
+\lambda\Phi$&$\begin{array}{l}C^1_1, \ C^1_2, \ K_3-P_3-2\lambda t \end{array}$&$\textsf{s}_{5,16}$\\
17& \vspace{2mm} $-\mu\Theta-\frac{\lambda^2}2\Theta^2+\nu\ln(\tilde
r)$& $\begin{array}{l}C^1_1,\ C^1_2,\ P_3,\
D+\nu t\end{array}  $&$\textsf{s}_{6,229}$\\
18& \vspace{2mm} $-\mu\Theta-\frac{\lambda^2}2\Theta^2$&
$\begin{array}{l}C^1_1,\ C^1_2,\ P_3,\
D, \ K_3\end{array}  $&$\textsf{s}_{4,7}\oplus$ sl(2,R)\\
\hline
\hline
\end{tabular}

 There is a big variety of potentials corresponding to
PDMs given by equations (\ref{m1}) and (\ref{m4}). The corresponding
classification results are arranged in Tables 1--5.

Equations (\ref{seq}) with any $m$ and $V$ admit symmetry operator
(\ref{P0}).
 Additional
symmetries are indicated in the third  columns of the tables where
$M_{ab},\ P_a, \ L_a$,  and $D$ are operators defined in (\ref{QQ})
and (\ref{iso}), $G(\cdot)$ are arbitrary functions of the arguments
fixed in brackets, and small Greek letters denote arbitrary
parameters.

\begin{center}Table 2\\
Systems with  power inverse PDM function  $f=\tilde r^2e^{\sigma
\Theta},\quad \sigma\neq0,\ \ \omega\neq 0$. The invariance algebras
include generator (\ref{P0}) and can be extended by adding the unit
operator. Potentials  presented in Items 6--10 and 11--15 can be
obtained from the ones presented in Items 1--5 using transformations
(\ref{et9}) and (\ref{et10}) correspondingly. Alternatively,
parameter $\omega$ can be reduced to unity applying transformation
(\ref{et03}).
\end{center}
\begin{tabular}{r l l c}
\hline No&Potential $V$&Symmetries&$\begin{array}{c}\text{Invariance}\\\text{algebras}\end{array}$
\\
\hline&&\\
1 & \vspace{2mm}
 $G(\tilde
r,x_3)e^{\sigma\Theta}$&$\begin{array}{l} Q^1_1=\ri t^2\p_t
+\frac{2t}\sigma L_3+\frac2{\sigma^2}e^{-\sigma\Theta}, \\
Q^1_2=\frac1\sigma L_3 +\ri
t\p_t\end{array}$&sl(2,R) \\
2& \vspace{2mm} $G(\tilde r)e^{\sigma\Theta}$&$\begin{array}{l}
Q^1_1, \
Q^1_2,\ P_3\end{array}$&sl(2,R)$\oplus$ $\textsf{n}_{1,1}$ \\
3& \vspace{2mm} $G(\frac{\tilde
r}r)e^{\sigma\Theta}$&$\begin{array}{l} Q^1_1, \
Q^1_2,\ D\end{array}$ &sl(2,R)$\oplus$ $\textsf{n}_{1,1}$\\
4& \vspace{2mm} $G(\frac{r^2+1}{\tilde
r})e^{\sigma\Theta}$&$\begin{array}{l} Q^1_1, \
Q^1_2,\ K_3-P_3\end{array}$ &sl(2,R)$\oplus$ $\textsf{n}_{1,1}$\\
5& \vspace{2mm} $\kappa e^{\sigma\Theta}$&$\begin{array}{l} Q^1_1,
\ Q^1_2, \ P_3, \ D,\ K_3\end{array}$ &sl(2,R)$  \oplus $ sl(2,R)\\
6& \vspace{2mm} $G(\tilde r,x_3)e^{\sigma\Theta}+\frac{\omega^2}2
e^{-\sigma\Theta}$& $\begin{array}{l} N^1_1=\omega\cos
(\omega\sigma t) L_3\\\vspace{2mm}-\sin(\omega\sigma t)\left(\ri
\p_t- \omega^2e^{-\sigma\Theta}\right) ,\\N^1_2=\omega\sin
(\omega\sigma t) L_3\\+\cos(\omega\sigma t)\left(\ri \p_t-
\omega^2e^{-\sigma\Theta}\right)
\end{array}$&so(1,2)\\
7& \vspace{2mm} $G(\tilde r)e^{\sigma\Theta}+\frac{\omega^2}2
e^{-\sigma\Theta}$& $N^1_1,\
N^1_2,\ P_3$&so(1,2)$\oplus\ $$\textsf{n}_{1,1}$ \\
8& \vspace{2mm} $G(\frac{\tilde
r}r)e^{\sigma\Theta}+\frac{\omega^2}2 e^{-\sigma\Theta}$&
$\begin{array}{l}N^1_1,\
N^1_2,\ D\end{array}$&so(1,2)$\oplus\ $$\textsf{n}_{1,1}$\\
9& \vspace{2mm} $G(\frac{r^2+1}{\tilde
r})e^{\sigma\Theta}+\frac{\omega^2}2 e^{-\sigma\Theta}$&
$\begin{array}{l}N^1_1,\
N^1_2,\ K_3-P_3\end{array}$&so(1,2)$\oplus\ $$\textsf{n}_{1,1}$ \\
10& \vspace{2mm} $\kappa e^{\sigma\Theta}+\frac{\omega^2}2
e^{-\sigma\Theta}$&
$\begin{array}{l}N^1_1, \ N^1_2,  P_3, \ D,\ K_3\end{array}$ &so(1,2)$  \oplus $sl(2,R)\\
11& \vspace{2mm} $G(\tilde r,x_3)e^{\sigma\Theta} -\frac{\omega^2}2
e^{-\sigma\Theta}$& $\begin{array}{l}S^1_1=\omega\cosh
(\omega\sigma t) L_3\\-\sinh(\omega\sigma t) \left(\ri \p_t+
\omega^2e^{-\sigma\Theta}\right),\vspace{2mm}\\
S^1_2=\omega\sinh (\omega\sigma t) L_3
\\ -\cosh(\omega\sigma t)\left(\ri
\p_t+ \omega^2e^{-\sigma\Theta}\right), \\
\end{array}$& so(1,2)\\
12& \vspace{2mm} $G(\tilde r)e^{\sigma\Theta}-\frac{\omega^2}2
e^{-\sigma\Theta}$& $\begin{array}{l}S^1_1, \
S^1_2,\ P_3\end{array}$&so(1,2)$\oplus\ $$\textsf{n}_{1,1}$ \\
13& \vspace{2mm} $G(\frac{\tilde
r}r)e^{\sigma\Theta}-\frac{\omega^2}2 e^{-\sigma\Theta}$&
$\begin{array}{l}S^1_1, \
S^1_2,\ D\end{array}$ &so(1,2)$  \oplus $ $\textsf{n}_{1,1}$\\
14& \vspace{2mm} $G(\frac{r^2+1}{\tilde
r})e^{\sigma\Theta}-\frac{\omega^2}2 e^{-\sigma\Theta}$&
$\begin{array}{l}S^1_1, \
S^1_2,\ K_3-P_3\end{array}$&so(1,2)$\oplus\ $$\textsf{n}_{1,1}$ \\
15& \vspace{2mm} $\kappa e^{\sigma\Theta}-\frac{\omega^2}2
e^{-\sigma\Theta}$&
$\begin{array}{l}S^1_1,\ S^1_2,\  P_3, \ D,\ K_3\end{array}$&so(1,2)$  \oplus$sl(2,R)\\

\hline
\hline
\end{tabular}

\vspace{2mm}

\newpage

\begin{center}Table 3\\
Systems with  power inverse PDM functions $f= r^2=x_1^2+x_2^2+x_3^2$
(Items 1-9) and $f= r^{2+\sigma},\ \sigma\neq0,\ \pm1$ (Items
10--18). Parameter $\mu$ can be reduced to zero
 using transformation inverse to (\ref{et5}) for Items 1--3 and
 transformation (\ref{et04}) for Items 4--9 . Potentials and symmetries presented in Items 13, 14, 15 and 16, 17, 18 can be obtained from the ones presented in Items 10, 11, 12 using transformations (\ref{et11}) and
 (\ref{et12}).  In addition,
parameter $\omega$ can be reduced to unity applying transformation
(\ref{et06}).\end{center}

\begin{tabular}{r l l c}
\hline No&Potential $V$&Symmetries&$\begin{array}{c}\text{Invariance}\\\text{algebras}\end{array}$
\\
\hline 1& \vspace{1mm} $G(\Theta,\frac{\tilde
r}r)+\mu\ln(r)$&$\begin{array}{l}A^2_1=tD +\frac12\mu
t^2-\ln(r), \\  A^2_2=D+\mu t\end{array}$&$\textsf{n}_{4,1}$\\
2&
\vspace{2mm}
$G(\frac{\tilde
r}r)+\mu\ln(r)+\kappa\Theta$&$\begin{array}{l}A^2_1, \ A^2_2, \ L_3+\kappa t\end{array}$&$\textsf{n}_{5,4}$\\
3&\vspace{1mm}$\mu\ln(r)$&$\begin{array}{l}A^2_1,\  A^2_2,
 \  L_1, \ L_2, \ L_3 \end{array}$&$\textsf{n}_{4,1}$$\oplus\  $so(3)\\
4&\vspace{1mm}$G(\Theta,\frac{\tilde
r}r)+\mu\ln(r)+{\frac{\lambda^2}2}\ln(r)^2 $&$\begin{array}{l}
B^2_1= \sin(\lambda t)D\\- \cos(\lambda
t)({\lambda}\ln(r)+\frac{\mu}\lambda),\\B^2_2=\cos(\lambda t)D\\+
\sin(\lambda
t)({\lambda}\ln(r)+\frac{\mu}\lambda)\end{array}$&$\textsf{n}_{4,7}$\\
5&\vspace{1mm}$G(\frac{\tilde r}r)+\kappa\Theta+
{\frac{\lambda^2}2}\ln(r)^2+\mu\ln(r)$&$\begin{array}{l}B^2_1, \
B^2_2, \ L_3+\kappa t\end{array}$&$\textsf{s}_{5,16}$\\
6&\vspace{1mm}$\mu\ln(r)+
{\frac{\lambda^2}2}\ln(r)^2$&$\begin{array}{l}B^2_1,\ B^2_2, \
 L_1, \ L_2, \ L_3 \end{array}$&$\textsf{n}_{4,7}$$\oplus\ $so(3)\\
7&\vspace{1mm}$G(\Theta,\frac{\tilde
r}r)-\mu\ln(r)-{\frac{\lambda^2}2}\ln(r)^2$&$\begin{array}{l}C^2_1=\cosh(\lambda
t)D\\+ \sinh(\lambda
t)({\lambda}\ln(r)+\frac{\mu}\lambda), \\
C^2_2= \sinh(\lambda t)D\\+ \cosh(\lambda
t)({\lambda}\ln(r)+\frac{\mu}\lambda)\end{array}$&$\textsf{n}_{4,6}$\\
8&\vspace{1mm}$G(\frac{\tilde r}r)+\kappa\Theta-
{\frac{\lambda^2}2}\ln(r)^2-\mu\ln(r)$&$\begin{array}{l}C^2_1, \
C^2_2, \ L_3+\kappa t\end{array}$&$\textsf{s}_{5,15}$\\
9&\vspace{3mm}$-\mu\ln(r)-
{\frac{\lambda^2}2}\ln(r)^2$&$\begin{array}{l}C^2_1,\ C^2_2,
\   L_1, \ L_2, \ L_3 \end{array}$&$\textsf{n}_{4,6}$$\oplus\ $so(3)\\
10&\vspace{1mm}$r^{\sigma}G(\Theta,\frac{\tilde r}r)$&
$\begin{array}{l}Q^2_1=\ri t^2\p_t+\frac{2t}\sigma D+
\frac{2}{\sigma^2 r^{\sigma}}, \\ Q^2_2=\frac1\sigma D+\ri t\p_t
\end{array}$&sl(2,R)\\
11&\vspace{1mm}$r^{\sigma}G(\frac{\tilde r}r)$&
$\begin{array}{l}Q^2_1,\ Q^2_2,\ L_3
\end{array}$&sl(2,R)$\oplus\ $$\textsf{n}_{1,1}$\\
12&\vspace{1mm}$\kappa r^{\sigma}$& $\begin{array}{l}Q^2_1,\ Q^2_2,\
L_1, \ L_2, \ L_3
\end{array}$&sl(2,R)$\oplus\ $so(3)\\
13&\vspace{1mm}$r^{\sigma}G(\Theta, \frac{\tilde
r}r)+\frac{\omega^2}2r^{-\sigma}
$&$\begin{array}{l}N^2_1=\omega\cos(\omega\sigma
t)D\\+\sin(\omega\sigma t)(\ri \p_t-{\omega^2} r^{-\sigma})
,\\ N^2_2=\omega\sin(\omega\sigma t)D\\-\cos(\omega\sigma t)(\ri
\p_t-{\omega^2} r^{-\sigma})
 \end{array}\ $&so(1,2)\\
14&\vspace{1mm}$r^{\sigma}G( \frac{\tilde
r}r)+\frac{\omega^2}2r^{-\sigma}
$&$\begin{array}{l}N^2_1, \ N^2_2, \ L_3\end{array}$&so(1,2) $\oplus\ $$\textsf{n}_{1,1}$\\
15\vspace{1mm}&$\kappa r^{\sigma}+
\frac{\omega^2}2r^{-\sigma} $&$\begin{array}{l}N^2_1,\
N^2_2,\
 L_1, \ L_2, \ L_3 \end{array}$&so(1,2)$\oplus\ $so(3)\\
16&\vspace{1mm}$r^{\sigma}G(\Theta, \frac{\tilde
r}r)-\frac{\omega^2}2r^{-\sigma}$ &$\begin{array}{l}
S^2_2=\omega\cosh(\omega\sigma t)D\\+ \sinh(\omega\sigma
t)(\ri\p_t+{\omega^2}r^{-\sigma})
,\\S^2_1=\omega\sinh(\omega\sigma t)D\\+\cosh(\omega\sigma t)(\ri
\p_t+{\omega^2}r^{-\sigma})
\end{array}$&so(1,2)\\
17&\vspace{1mm}$r^{\sigma}G( \frac{\tilde
r}r)-\frac{\omega^2}2r^{-\sigma}$
&$\begin{array}{l}S^2_1, \ S^2_2, \ L_3\end{array}$&so(1,2)$\oplus\ $$\textsf{n}_{1,1}$\\
18& \vspace{1mm} $\kappa r^{\sigma}-
\frac{\omega^2}2r^{-\sigma}$
&$\begin{array}{l}S^2_1,\ S^2_2, \  L_1, \ L_2, \ L_3 \end{array}$&so(1,2)$\oplus\ $so(3)\\
\hline\hline\end{tabular}

\vspace{3mm}

Symmetries  collected in Tables 1 and 2 are of type 1. The
corresponding vectors $\xi^k$ have the type presented in
(\ref{e71}), (\ref{q7})--(\ref{q9}). The related symmetry algebras
are fixed in the fourth columns, where
 $\textsf{n}_{a,b}$ and $\textsf{s}_{a,b}$ are nilpotent and solvable Lie
 algebras of dimension $a$.   We use the notations presented in \cite{snob}
 for low dimension  Lie algebras. A discussion of symmetry
 algebras which are admitted by equation (\ref{seq}) can be found
in Section 7.

In Table 3 we present symmetries of type (\ref{so}), (\ref{e72}).
The maximally extended
 symmetry algebras are seven dimensional, see Items 3, 6 and 9 there.
 However, these algebras and all other algebras presented in Items
 1, 2, 4, 5, 7, 8  ultimately include the
 unit operator which is accepted by any of the
 considered equations. The invariance algebras presented in Items 10--18
can be extended by adding the unit operator. Algebra
$\textsf{n}_{1,1}$ includes the only basis element $L_3$.

 Notice that the low dimension algebras with dimension $d\leq5$ and a certain class of the algebras with dimension 6 had been classified by
 Mubarakzianov \cite{mur}, see also a more contemporary
 and accessible papers \cite{bas}, \cite{boy1} and \cite{boy2}.

In Table 4 some exotic PDMs are specified. In particular, there are
two systems which admit eight  dimensional invariance algebras, see
Items 6 and 7. However, the related symmetries are time independent
and commute with Hamiltonians. In other words, they belong to
integrals of motion for the stationary Schr\"odinger equation which
have been  classified in paper \cite{NZ}.\footnote{The list of integrals
of motion presented in \cite{NZ} includes two extra cases which can
be omitted without loss of generality. For the reduced list of
inequivalent integrals of motion see the latest version of  preprint
arXiv:1412.4332   }
\begin{center}Table 4\\
Systems with some particular mass functions.  The invariance
algebras include the presented symmetries together with generator
(\ref{P0}) and the unit operator. Parameter $\mu$ can be reduced to
zero using transformation inverse to (\ref{et07}).
\end{center}
\begin{tabular}{ccllc}
\hline No&$\begin{array}{c}\texttt{Inverse}\\ \texttt{mass} \ f
\end{array}$&Potential $V$&Symmetries&$\begin{array}{c}\text{Invariance}\\
\text{algebras}\end{array}$
\\
\hline &&&&\\

1&\vspace{2mm}$\tilde r^3$&$ \kappa x_3+\lambda\tilde r$&$ P_3+\kappa t,
\ D+\ri t\p_t, \ L_3$&$\textsf{s}_{4,6}$$\oplus\ $$\textsf{n}_{1,1}$\\
2&\vspace{2mm}$x_3^2$&$\mu\ln(x_3)$&$P_1,\ P_2,\ L_3,\ D+\mu
t$&$\textsf{s}_{5,3}\oplus\textsf{n}_{1,1}$
\\
3&\vspace{2mm}$x_1^3$&$\lambda x_1+\kappa x_3$&$P_3+\kappa t, \ P_2,  \
D+\ri t\p_t$&$\textsf{s}_{5,17}$\\
4&\vspace{2mm}$x_3^{\sigma+2}$&$\kappa x_3^\sigma$&$P_1,\ P_2,\ L_3, \
D+\ri\sigma t\p_t, \ \sigma\neq
0,1,-2$&$\textsf{s}_{4,3}\oplus\textsf{n}_{1,1}\oplus\textsf{n}_{1,1}$\\
5&\vspace{2mm}$\tilde r^{\sigma+2}e^{\lambda\Theta}$&$\kappa \tilde
r^{\sigma}e^{\lambda\Theta}$&$L_3+\ri\lambda t\p_t,\ P_3,\
D+\ri\sigma t\p_t,\ \sigma\neq0$&$\textsf{s}_{2,1}\oplus\textsf{s}_{2,1}\oplus\textsf{n}_{1,1}$\\
    6&$(r^2+1)^2
         $&\vspace{2mm}$-3 r^2$&$M_{41}, M_{42}, M_{43}, M_{21}, M_{31}, M_{32}$&so(4)$\oplus\textsf{n}_{1,1}\oplus\textsf{n}_{1,1}$ \\
  7&\vspace{2mm}$(r^2-1)^2$&$-3 r^2$&$ M_{01}, M_{02}, M_{03}, M_{21}, M_{31},
   M_{32}$&so(1,3)$\oplus\textsf{n}_{1,1}\oplus\textsf{n}_{1,1}$ \\
   \hline\hline
   \end{tabular}

   \vspace{2mm}
All the invariance algebras presented in Items 1--5 of Table 4 are
solvable. Like in the previous tables we use notations
$\textsf{s}_{a,b}$ for solvable algebra of dimension $a$ and
classification number $b$, see \cite{snob}.

One dimension algebra $\textsf{n}_{1,1}$ in Items 1, 2 is spanned on
$L_3$ and the unit operator correspondingly. In Item 4 the basis
elements of $\textsf{s}_{1,1}\oplus\textsf{s}_{1,1}$ are $L_3$ and
the unit operator. In Items 6 and 7 the symbol
$\textsf{s}_{1,1}\oplus\textsf{s}_{1,1}$ denotes the direct sum of
one dimensional algebras spanned on $P_0$ and the unit operator.

Symmetries of type (\ref{so}), (\ref{e73}) are collected in Table 5.
The corresponding masses are exponentials in the only spatial
variable $x_3$. The invariance algebras again have dimensions 3, 4,
5 or 6. These algebras are either simple of form a direct sum of a
simple and solvable algebra, and can be extended by the unit
operator. By e(2)$\sim\textsf{s}_{3,3}$ we denote the Euclidean
algebra in two dimensional space, whose generators are $P_1, P_2$
and $L_3$, while $\textsf{n}_{1,1}$ denotes the one--dimensional
algebras whose generators are presented in the last positions of
Column 3.

\begin{center}Table 5\\
Systems with  exponential inverse PDM function $f=e^{ x_3}$. Potentials and
symmetries presented in Items 5, 6, 7, 8 and 9, 10, 11, 12 can be obtained
from the ones presented in Items 1, 2, 3, 4 using transformations (\ref{et7})
 and (\ref{et8}). In addition,
parameter $\omega$ can be reduced to unity applying transformation
(\ref{et06}), and equivalence transformation (\ref{et6}) is
 available which generates multiplier $\sigma$ for $x_3$. \end{center}
\begin{tabular}{rllc}
\hline No&Potential $V$&Symmetries&$\begin{array}{c}\text{Invariance}\\
\text{algebras}\end{array}$
\\
\hline &&\\
1&\vspace{1mm}$e^{
x_3}G(x_1,x_2)$&$\begin{array}{l}Q^3_1=\ri
t^2\p_t+2tP_3+{2}e^{-x_3}, \\
Q^3_2=\ri
t\p_t+P_3\end{array}$&sl(2,R)\\
2&\vspace{1mm}$ e^{
x_3}G(x_2)$&$\begin{array}{l}Q^3_1, \ Q^3_2, \ P_1 \end{array}$&sl(2,R)$\oplus$ $\textsf{n}_{1,1}$\\
3&\vspace{1mm}$e^{x_3}G(\tilde r)$&$\begin{array}{l}Q^3_1, \
Q^3_2, \ L_{3}\end{array}$&sl(2,R)$\oplus$ $\textsf{n}_{1,1}$\\
4&\vspace{1mm}$\kappa e^{x_3}$&$\begin{array}{l}Q^3_1,\
Q^3_2, \ P_1, \ P_2,\
L_3\end{array}$&sl(2,R)$\oplus$ e(2)\\
5&\vspace{1mm}$e^{
x_3}G(x_1,x_2)+\frac{\omega^2}2e^{-
x_3}$&$\begin{array}{l}N^3_1= \sin(\omega t)\left(
\ri\p_t-\omega^2{\text e}^{-
x_3}\right)\\+\omega\cos(\omega
t)P_3,\\N^3_2=\cos(\omega t)\left(
\ri\p_t-\omega^2{\text e}^{-
x_3}\right)\\-\omega\sin(\omega
t)P_3\end{array}$&so(1,2)\\
6&\vspace{1mm}$e^{
x_3}G(x_1)+\frac{\omega^2}2e^{-
x_3}$&$\begin{array}{l}N^3_1, \ N^3_2, \ P_2\end{array}$&so(1,2)$\oplus$ $\textsf{n}_{1,1}$\\
7&\vspace{1mm}$e^{x_3}G(\tilde
r)+\frac{\omega^2}2e^{-
x_3}$&$\begin{array}{l}N^3_1, \ N^3_2, \ L_3\end{array}$&so(1,2)$\oplus$ $\textsf{n}_{1,1}$\\
8&\vspace{1mm}$\kappa e^{
x_3}+\frac{\omega^2}2e^{-
x_3}$&$\begin{array}{l}N^3_1,\ N^3_2,\ P_1,
\ P_2, \ L_3\end{array}$&so(1,2)$\oplus$ e(2)\\
9&\vspace{1mm}$e^{
x_3}G(x_1,x_2)-\frac{\omega^2}2e^{-
x_3}$&$\begin{array}{l}
S^3_1=\sinh(\omega t)(\ri\p_t+\omega^2{\text
e}^{-
x_3})\vspace{1mm}\\+\omega\cosh(\omega t)P_3,\vspace{1mm}\\\vspace{1mm}S^3_2= \cosh(\omega
t)(\ri\p_t+\omega^2{\text e}^{-
x_3})\\+\omega\sinh(\omega t)P_3\end{array}$&so(1,2)\\
10\vspace{1mm}&$e^{
x_3}G(x_1)-\frac{\omega^2}2e^{-
x_3}$&$\begin{array}{l}S^3_1, \ S^3_2, \ P_2\end{array}$&so(1,2)$\oplus\ $$\textsf{n}_{1,1}$\\
11\vspace{1mm}&$e^{x_3}G(\tilde
r)-\frac{\omega^2}2e^{-
x_3}$&$\begin{array}{l}S^3_1,
\ S^3_2, \ L_3\end{array}$&so(1,2)$\oplus\ $$\textsf{n}_{1,1}$\\
12&\vspace{1mm}$\kappa e^{
x_3}-\frac{\omega^2}2e^{-
x_3}$&$\begin{array}{l}S^3_1,\ S^3_2,
\  P_1,   \ P_2, \ L_3\end{array}$&so(1,2)$\oplus\ $e(2)\\
\hline\hline\end{tabular}

\vspace{3mm}

   In Table 6 rather generic systems are presented. They include
 potentials (and masses) defined up to arbitrary functions, but admit
 reduced numbers of symmetries belonging to class 3 specified in Section 5.  In Table 6 the symbols $F(\cdot)$ and $G(\cdot)$ denote arbitrary functions of
arguments
 fixed in brackets,
  $D_1F$ and $D_2F$  are the derivations of function
  $F=F\left(\frac{r^2+1}{\tilde r},\nu\Phi+\Theta\right)$ with respect to
  the first and second argument  correspondingly,  $F'$ is the derivation
  of function $F=F\left(\frac{r^2\pm1}{\tilde r}\right) $ with respect to
  its combined argument $\frac{r^2\pm1}{\tilde r}$. In addition, to save a
  room  we denote  $\Phi=\arctan\left(\frac{r^2-1}{2x_3}\right)$ and
    $\Psi= \frac{3\nu x_3\tilde r^2}{(r^2+1)^2-4\tilde r^2}.$

\begin{center}Table 6.\\Systems with  masses and potentials defined up to
arbitrary functions\end{center}
\begin{tabular}{rlll}
\hline No&Inverse mass $f$&Potential&Symmetries
\\
\hline\\
1&$F(x_1,x_2)e^{\lambda x_3}$ \vspace{2mm} &$\begin{array}{l}G(x_1 ,x_2)
e^{\lambda x_3}\end{array}$&$P_3+\ri\lambda t\p t$ \\
2&$F(x_1,x_2)$
\vspace{2mm}
&$G(x_1 ,x_2)+\lambda x_3$&$P_3+\lambda t$\\

 3&$F(\tilde r,x_3)e^{\sigma\Theta}$
\vspace{2mm}
&$G(\tilde r,x_3)
e^{\sigma\Theta}$&$L_3+\ri\sigma t\p_t$\\
4&$F(\tilde r,x_3)$
\vspace{2mm}
&$\begin{array}{l}G(\tilde r,x_3)\ \
+\lambda \Theta\end{array}$&$L_3+\lambda t$\\
5&$r^{\sigma+2}F(\frac{\tilde r}{ r}, r^\lambda e^ {-\Theta})$
\vspace{2mm} &$r^\sigma G(\frac{r}{\tilde r},r^\lambda e^
{-\Theta})$&$\lambda L_3 +D+\ri\sigma t\p_t$\\
6 &$r^{2}F(\frac{\tilde  r}{r}, r^\lambda e^{ -\Theta})$
\vspace{2mm} &$\begin{array}{l}G(\frac{r}{\tilde r},r^\lambda e^{
-\Theta})\\+{\sigma(\lambda\Theta+\ln(r))}\end{array}$&$\begin{array}{l}
\lambda L_3 +D\\+\sigma(1+\lambda^2) t\end{array}$\\
7 &$F(\tilde r, x_3-\Theta)e^{\sigma \Theta}$ \vspace{2mm} &$
G(\tilde r, x_3-\Theta)e^{\sigma \Theta}$&$P_3+L_3+\ri\sigma t\p_t$\\
8 &$F(\tilde r, x_3-\Theta)$&$G(\tilde r, x_3-\Theta)+\lambda\Theta$
\vspace{2mm}
&$P_3+L_3+\lambda t$\\
9&$ \tilde r^2F\left(\frac{r^2+1}{\tilde
r},\lambda\Phi+\Theta\right)e^{\sigma \Theta}$ \vspace{2mm} &
$\begin{array}{l}e^{\sigma \Theta}\left(\Psi D_2F-\frac{3\tilde
r}2D_1F\right.\\\left.+G\left(\frac{r^2+1}{\tilde
r},\lambda\Phi+\Theta\right) \right)\end{array}$&$M_{43}+\lambda
L_3+\ri\sigma t\p_t$,
\\
10&$ \tilde r^2F\left(\frac{r^2+1}{\tilde
r},\lambda\Phi+\Theta\right)$ \vspace{2mm} &
$ \begin{array}{l}G\left(\frac{r^2+1}{\tilde r},\lambda\Phi+\Theta\right)+\sigma \Theta \\-\frac{3\tilde r}2D_1F+\Psi D_2F \end{array}$&$M_{43}+\lambda L_3+\ri\lambda\sigma t, $\\

11& $F(\tilde r)e^{\sigma\Theta+\lambda x_3}$ \vspace{2mm} &
$G(\tilde r)e^{\sigma\Theta+\lambda
x_3}$&$\begin{array}{l}L_3+\sigma t
\p_t,\ P_3+\ri\lambda t\p_t\end{array}$\\
12&$F(\tilde r)$
\vspace{2mm}
&$G(\tilde r)+\sigma\Theta+\lambda x_3$&$L_3+\sigma t,\ P_3+\lambda t$\\
13&$\tilde r^{\sigma+2}e^{\lambda\Theta}F(\frac{\tilde r}r)$
\vspace{2mm} & $\tilde r^{\sigma}e^{\mu \Theta}G(\frac{x_3}{\tilde
r})$&$\begin{array}{l}
 D+\ri\sigma t\p_t,\
  L_3+\ri\lambda t\p_t\end{array}$\\
14&$\tilde r^{2}F(\frac{\tilde r}r)$ \vspace{2mm} & $G(\frac{\tilde
r}{ r})+\sigma \Theta+\lambda\ln(r)$&$\begin{array}{l}
 D+\lambda t,\
  L_3+\sigma t\end{array}$
   \\
  15&$e^{\sigma x_1}F(x_3)$
   \vspace{2mm}
  &$e^{\sigma x_1}G(x_3)$&$P_1+\ri\sigma t\p_t,\ P_2$\\
16&$F(x_3)$&$G(x_3)+\lambda x_2$
 \vspace{2mm}
&$P_1,\ P_2+\lambda t$\\

17&$e^{\sigma\Theta-\lambda\Phi}\tilde r^2 F(\frac{r^2+1}{\tilde
r})$
 \vspace{2mm}
&$ e^{\sigma\Theta-\lambda\Phi}\left(\frac32\tilde rF'+
G(\frac{r^2+1}{\tilde r})\right)$&$M_{43}+\ri\lambda t\p_t ,\ L_3+\ri\sigma t\p_t $  \\
18&$\tilde r^{2} F(\frac{r^2+1}{\tilde r})$ &$\begin{array}{l}
\frac32\tilde rF'+ G(\frac{r^2+1}{\tilde
r})\\+\sigma\Theta-\lambda\Phi\end{array}$
 \vspace{2mm}
&$M_{43}+\lambda t,\ L_3+\sigma t$  \\
19&$\tilde r^{\sigma+2}F(\tilde r^\nu e^{-\Theta})$&$\tilde
r^{\sigma}G(\tilde r^\nu e^{-\Theta})$
 \vspace{2mm}
&$D+\ri\sigma t\p_t+\nu L_3,
P_3$\\
20&$\tilde r^{3}F(\tilde r^\nu e^{-\Theta})$
 \vspace{2mm}
&$\tilde rG(\tilde r^\nu e^{-\Theta})+\sigma x_3$&$D+\ri
t\p_t+\nu L_3,
P_3+\sigma t$\\
21&$\tilde r^{2}F(\tilde r^{\nu}\text{e}^{-\Theta})$
 \vspace{2mm}
&$\sigma\ln(\tilde r)+G(\tilde
r^{\nu}\text{e}^{-\Theta})$&$D+\nu L_3
+\sigma t,\ P_3$\\
22&$F(x_3)$&$G(x_3)$
 \vspace{2mm}
&$P_1,\ P_2,\ L_3$\\
  23&$F(r^2)$&$G(r^2)$
   \vspace{2mm}
  &$L_1,\ L_{2},\ L_{3}$\\
    24&$x_3^{2}F\left(\frac{r^2-1}{x_3}\right)$&$\frac32x_3 F'
  +G\left(\frac{r^2-1}{x_3}\right)$
   \vspace{2mm}

  &$M_{01},\ M_{02},\ M_{12}$\\
    \hline\hline
\end{tabular}

 We indicate   symmetries additional to $P_0$ and $I$ in the third column.
  For the systems presented in
 Items 1, 3, 5, 7, 9 and 2, 4, 6, 8, 10 the corresponding symmetry algebras
 are $\textsf{s}_{2,1}\oplus \textsf{n}_{1,1}$ and $\textsf{s}_{3,1}$
 correspondingly. In Items 11, 13, 15, 17, 19 and 21 we have symmetry
 algebras isomorphic to  $\textsf{s}_{2,1}\oplus \textsf{n}_{1,1}
 \oplus \textsf{n}_{1,1}$ if at least one of parameters $\lambda$ or
 $\sigma$ is nontrivial. If all these parameters are equal to zero, these
  symmetry algebras (and algebras admitted by systems presented in Items 12, 14, 16, 18 and 20) are degenerated to  the direct sums of four one dimensional algebras.

  In Items 12, 14, 16, 18 and 20 we have symmetry algebras isomorphic to $\textsf{s}_{3,1}\oplus \textsf{n}_{1,1} $ if at least one of parameters $\lambda$ or
 $\sigma$ is nontrivial. Finally, the systems presented in Items 22, 23 and 24 admit the direct sums of symmetry algebras $e(2)\oplus\textsf{n}_{1,1}$,
so(3)$\oplus\textsf{n}_{1,1}$ and so(1,2)$\oplus\textsf{n}_{1,1}$ correspondingly. A
short discussion of the invariance algebras is presented in the
following section.

Thus we find all 3d Hamiltonians (\ref{A11}) with PDM which correspond to equations
 (\ref{seq}) with nonequivalent symmetries.
The presented list of equations (\ref{seq}) is completed up to
equivalence transformations belonging to group C(3). The additional
equivalence transformations are indicated in the tables headings. In all tables parameters $\mu$ and $\omega$ can be
reduced to zero using additional equivalence transformations
(\ref{et02})--(\ref{et12}), like the isotropic harmonic oscillator
can be reduced to the free particle Schr\"odinger equation. We keep
these parameters in the classification tables for the readers
convenience.

 The number of such equations
 is rather extended. Namely, we specify 22 classes of equations defined up
 to arbitrary parameters and 70 classes of equations defined up to arbitrary
 functions, see Tables 1--6. In addition, there are two systems with the fixed mass and potential terms presented in Items 6 and 7 of Table 4.

We did not consider  equations with constant mass terms,
 since they had been classified long time ago.
 However, some of such equations appears as particular cases of our
 analysis, and the comparison of these cases with well known results leads
 to a rather non-excepted conclusion: the classification results
 presented in \cite{Boy} are incomplete. A discussion of this point is presented
 in   Section 9.

\section{Algebras of symmetry operators}
The complete sets of Lie symmetries of a given partial differential  equation should have a structure of a Lie algebra. In particular, it is the case for symmetries presented in
 Tables 1--6. Let us describe these structures explicitly.

First we present commutation relations between essentially time dependent symmetry operators and generator $P_0$. For symmetries presented in Table
 1 and the first half of Table 3 we obtain:
\begin{gather}\la{CR1}[A^a_1,A^a_2]=-\ri I,\quad [P_0,A^a_2]=\ri\mu I,
\quad  [P_0,A^a_1]=\ri A^a_2;\\\la{CR2}[B^a_1,B^a_2]=-\ri\lambda I,\quad
[P_0,B^a_1]=\ri\lambda B^a_2,\quad [P_0,B^a_2]=-\ri\lambda
B^a_1;\\\la{CR3}[C^a_1,C^a_2]=-\ri\lambda I,\quad [P_0,C^a_1]=\ri\lambda
C^a_2,\quad [P_0,C^a_2]=\ri\lambda C^a_1 \end{gather} where $I$ is
the unit operator, $a=1, 2$ and no sum with respect to the repeating
indices.

Thus for systems indicated in Items 1, 7, 13 of Table 1 and Items 1,
4, 7 of Table 3 we have four dimensional symmetry algebras, which
are solvable and include the unit operator as a central element. In
accordance with  \cite{snob}, we use the notations
$\textsf{n}_{4,1}$ for algebras (\ref{CR1}) and $\textsf{n}_{4,7}$
for algebras (\ref{CR2}) and (\ref{CR3}).

The additional symmetries which are presented in the remaining items of the mentioned tables   amend the mentioned algebras  to five dimensional algebras $\textsf{n}_{5,4}$ or $\textsf{s}_{5,16}$.  In addition, in Items 6, 12 and 18 of Table 1 the direct sums of solvable five dimensional algebras with the simple algebra sl(2,R) are indicated. The latter is a linear span of basis elements $K_3, P_3$ and $D$.

For symmetries represented in Tables 2, 4 and Items 10--18 of Table 3
the following commutation relations hold:
\begin{gather}\la{cr1}[Q^k_1,Q^k_2]=-\ri Q^k_1,\quad [P_0,Q^k_1]=2\ri Q^k_2,\quad [P_0,Q^k_2]=-\ri P_0,\\\la{cr2}
[N^k_1,N^k_2]=-2\ri\omega P_0,\quad [P_0,N^k_1]=2\ri\omega
N^k_2,\quad [P_0,N^k_2]=-2\ri\omega
N^k_1,\\\la{cr3}[S^k_1,S^k_2]=\ri\omega P_0,\quad
[P_0,S^k_1]=\ri\omega S^k_2,\quad [P_0,S^k_2]=\ri\omega
S^k_1\end{gather} where $k=1,2,3$.

The corresponding Lie algebras are three dimensional and simple. Commutation relations (\ref{cr1}) specify  algebra  sl(2,R), while relations (\ref{cr2}) and (\ref{cr3}) define  algebra so(1,2) up to normalization of basis elements.  Notice that algebras sl(2,R) and so(1,2) are isomorphic.

 The additional symmetries presented in Tables 2, 3 and 5 commute with $Q^k_a, N^k_a$ and $S^k_a$ and satisfy the following commutation relations  between themselves:
\begin{gather}\la{cr4} [D,P_3]=\ri P_3,\quad [D,K_3]=-\ri K_3,\quad [P_3,K_3]=2\ri D,\\\la{cr5}[L_1,L_2]=\ri L_3, \quad [L_3,L_1]=\ri L_2,\quad [L_2,L_3]=\ri L_1,\\\la{cr6}[P_1,P_2]=0,\quad [P_1,L_3]=-\ri P_2,\quad [P_2,L_3]=\ri P_1.\end{gather}

Relations (\ref{cr4}), (\ref{cr5}) and (\ref{cr6}) specify algebras sl(2,R),
 so(3) and e(2) correspondingly.

\section{Equivalence relations for ambiguity parameters}
The subjects of our classification are equations (\ref{seq})
including PDM Hamiltonians of special form (\ref{A11}). In other
words, we restrict ourselves to the a priori fixed values of the
ambiguity parameters \begin{gather}\la{APa}\sigma=\gamma=-\frac12,\
\beta=0.\end{gather} Nevertheless the obtained results can be
 easily reformulated for arbitrary values of these parameters. To this
 effect it is sufficient to change the potentials presented in the tables
 to the corresponding effective potentials $\hat V$ given by relation (\ref{simsim}).

 Let us note that the ordering ambiguity for the kinetic energy term
 is an interesting problem which attracted attention of numerous
 researchers. The ordering (\ref{A11}) was proposed in \cite{ben} to ensure the current
 conservation. The ordering with $\alpha=\gamma=-\frac12$ was applied
 in \cite{Zhu} for reformulating the
connection rule problem on the two sides of an heterojunction. The
same ordering was derived in \cite{Cav} as the nonrelativistic limit
of the Dirac Hamiltonian with PDM, and in \cite{yung} via path
integral evaluation. There are other  ordering versions suitable for
particular  physical models: $\alpha=\gamma$ \cite{mora},
$\beta=\alpha=0, \gamma=-1$ \cite{gor}, $\gamma=0,
\alpha=\beta=-\frac12$ \cite{li}, $\alpha=\gamma=-\frac14,
\beta=-\frac12$ \cite{mustafa}. All the corresponding Hamiltonians
(\ref{A1}) are equivalent to (\ref{A11}) with effective potential
(\ref{simsim}) which in general differs from the initial potential
$V$.

Let us consider an abstract Hamiltonian (\ref{A1}) and transform it
to the form (\ref{A12}), i.e., make the change
\begin{gather}\la{1-2}\alpha\to\tilde\alpha, \quad \beta\to\tilde\beta,\quad \gamma\to\tilde \gamma, \quad \hat V\to V\end{gather}
were potentials $\hat V$ and $V$ are connected in the way given by
formula (\ref{simsim}).

We say Hamiltonian (\ref{A1}) is {\it ambiguity invariant}  if the transformed potential V is equal to the initial potential $\hat V$  up to a constant term: $\hat V=V+C$.
In accordance with (\ref{simsim}) it happens if inverse mass $f$ satisfies
the following condition:
\begin{gather}\la{simi}(\tilde\beta-\beta)ff_{aa}+
2(\alpha\gamma-\tilde\alpha\tilde\gamma){f_a f_a}=4Cf.\end{gather}

Equation (\ref{simi}) gives the necessary and sufficient conditions
of the complete equivalence of two generic Hamiltonian (\ref{A1})
and (\ref{A12}) with the same inverse mass functions $f=\frac1m$ and
potentials but different ambiguity parameters.  In particular, for
$\tilde\beta=-1$ and $\tilde \alpha=\tilde\gamma=0$ this condition
is reduced to the following form
\begin{gather}\la{simka}2 \alpha\gamma{f_a
f_a} +(\alpha+\gamma)f_{aa}f =4Cf\end{gather} while the second
Hamiltonian is reduced to the form given by equation (\ref{A11}).

For all the inverse masses $f$ represented in Tables 1--5 there are
multiparameter families of the ambiguity parameters satisfying
condition (\ref{simka}) with $C=0$. For the cases  included into
Tables 1, 2, 5 and position 2 of Table 4  the parameters
equivalent to (\ref{APa})
 have to satisfy the following condition
\begin{gather}\la{para}2\alpha\gamma+\alpha+\gamma=0 \end{gather}
which is a consequence of (\ref{simka}) with $C=0$.

Formula (\ref{para}) defines a family of the ambiguity parameters,
one of which, say, $\alpha$, can take arbitrary values different from $-\frac12$, whereas the
others are can be expressed via $\alpha$ in the following manner:
\begin{gather*}\gamma=-\frac\alpha{2\alpha+1},\quad
\beta=-\frac{2\alpha(\alpha+1)+1}{2\alpha+1},\quad
\alpha\neq-\frac12.\end{gather*}

Thus we  have a one parametric family of the ambiguity parameters
for which Hamiltonians (\ref{A1}) and (\ref{A12}) are ambiguity
equivalent provided their mass and potential terms are presented in
Tables 1 and 3. The same  is true
for the systems presented in Table 3, however the condition
(\ref{para}) should be changed by the following one:
\begin{gather*}2\alpha\gamma(\sigma+2)
+(\alpha+\gamma)(\sigma+3)=0\end{gather*} were $\sigma$ can take
both nonzero and zero values, the latter case corresponds to the
systems presented in Items 1--9.

For the systems presented in Table 1 and Items 10--18 of Table 3
there exist  a more general
 family of equivalent ambiguity parameters with $\beta=-1-\alpha-\gamma$ and
 arbitrary $\alpha$ and $\gamma$. In this case condition (\ref{para}) includes a
 non-trivial  $C$ and the corresponding Hamiltonians (\ref{A1}) and (\ref{A12})
 differ by a constant term, i.e.,  $V=\hat V+C$ which is not essential and
 can be removed by the change $\psi\to e^{-\ri C}\psi$
 in equation (\ref{A11}).

It is interesting to note that some systems presented in the
classification tables can be treated as purely kinematical. This
means that the corresponding Hamiltonians can be represented in
generic form (\ref{A1}) with suitable ambiguity parameters and {\it
trivial or constant potential $V$}.

First, it is the case when potentials do not include parameters,
which can play the role of coupling constants, see Items 6 and 7 of
Table 4. The corresponding Hamiltonians can be rewritten in the form
(\ref{A12}) with $\tilde\sigma=\tilde \gamma=-\frac12,\ \tilde
\beta=0$ and $\tilde V=0$. The same property have Hamiltonians which
include masses and potentials specified in Items 4 and 5 of this
table, Item 5 of Table 2, Item 12 of Table 3 and Item 4 of Table 5,
in spite of that they include coupling constant $\kappa$.

Moreover,
 all potential terms in Tables 2-5 including parameter $\kappa$ can be
 transferred into the kinetic part of the Hamiltonians with some special values
 of the ambiguity parameters. These values can be easily found
 solving the corresponding equation (\ref{simka}) where $C$ is changed by the
 potential term with the multiplier $-\kappa$. In
 particular, for the Hamiltonian whose  potential term is
 specified in Item 5 of Table 2 we obtain the following condition
 for the ambiguity parameters:
 \begin{gather}\la{kap}(\sigma+2)\left(2\alpha\gamma(\sigma+2)+
 (\alpha+\gamma)
 (\sigma+3)\right)=-4\kappa.\end{gather}

 All PDM Hamiltonians (\ref{A1}) with the ambiguity parameters satisfying
 (\ref{kap}) and trivial potential $\hat V=0$ can be rewritten in form (\ref{A11})
 with $V=\kappa r^\sigma$.

  \section{New symmetries of 3d Schr\"odinger equation with constant mass term}
It is generally accepted to think that symmetry classification of
Schr\"odinger equations with constant masses has been completed long
time ago in Boyer paper \cite{Boy}. Let us show that, in order to be
complete, the Boyer classification should be extended to include
some systems missing in \cite{Boy}.

We considered equations with essentially nonconstant masses and ignored the systems
with $f=Const.$ However, some of  systems with constant masses are presented implicitly
in the classification tables. Thus, the systems presented in Table 2
include PDMs defined up to arbitrary  parameter $\sigma\neq0$.
However, they are well defined for $\sigma=-1$, but the
corresponding mass functions used in Items 10--18 are reduced to
constants.

All PDMs  presented in Table 6 are defined up to arbitrary
functions, which in particular can be chosen as constants. In this
way, taking if necessary $\sigma=-2$, many classes of PDM
Schr\"odinger equations specified in  Table 6 can be reduced to
equations with constant mass terms.

Thus Tables 3 and 6 specify implicitly a number of constant
mass Schr\"odinger equations  which admit non-equivalent symmetries,
and it is possible to compare these results with the results of
Boyer classification \cite{Boy}.

Doing this, it is possible to see that there are at least three
cases missing in \cite{Boy}. Namely, setting in Items 4,  8 and 12
of Table 6 $F=1$ and substituting the corresponding expressions for
$f$ and $V$ into (\ref{A11}) we come to the following Hamiltonians:
\begin{gather}H=\frac{p^2}{2}+G(\tilde r, x_3)+
\sigma \Theta,\la{mi1}\\H=\frac{p^2}{2}+G(\tilde r,
\Theta-x_3)+\sigma \Theta, \la{mi2}\\ H=\frac{p^2}{2}+G(\tilde
r)+\sigma \Theta+\mu x_3. \la{mi3}\end{gather} Equation (\ref{seq})
with Hamiltonian (\ref{mi1}) admits the symmetry:
\begin{gather}\la{sy1}Q=\ri(x_2\p_1-x_1\p_2)+\sigma t,\end{gather}
while for Hamiltonian (\ref{mi2}) we have:
\begin{gather}\la{sy2}\tilde Q=Q-\ri\p_3.\end{gather}
Finally, equation (\ref{seq}) with Hamiltonian (\ref{mi3}) admits tree symmetry operators, namely,  (\ref{sy1}) and (\ref{sy3}):
\begin{gather}\la{sy3}\hat Q=\ri\p_3-\mu t,\quad G=\ri t\p_3+x_3.\end{gather}

Symmetries $Q,\ \tilde Q$ and $\hat Q$ are fixed in Table 6.
Symmetry $G$ is not present there since it appears only in the case
of a constant mass.

Equations (\ref{mi1})--(\ref{mi3}) are missing in Boyer
classification results, the same is true for symmetries of type
(\ref{sy1}). However, such symmetries do exist, and this fact can be
verified directly, calculating commutators of operators (\ref{sy1})
and (\ref{sy2}) with $L$ defined in (\ref{seq}). Such commutators
are identically equal to zero.

\section{A little on supersymmetry}
Some of the presented equations admit rather extended symmetries and so have good chances to be exactly solvable. In  many cases the extended symmetry guaranties the existence of solutions in separated variables, i.e., the equation is separable or even multiseparable.

In this section we discuss one more nice property of some of the presented equations. This property is called shape invariance, which can be considered as a reason for the exact solvability. Just shape invariant systems can be solved algebraically using tools of
SUSY quantum mechanics.
\subsection{Sharp invariance with oscillator potential}
Let us consider equation (\ref{seq}), (\ref{A11}) where $f$ and $V$ are
 functions fixed in Item 15 of Table 3:
\begin{gather}\la{equ5}\ri\frac{\p \psi}{\p t}=
\left(-\frac12\p_a r^{2\nu+2}\p_a+\kappa r^{2\nu}+
\frac{\omega^2}{r^{2\nu}}\right)\psi\end{gather}
where we denote $\sigma=2\nu$.

Equation (\ref{equ5}) admits rather extended symmetries being invariant w.r.t. six parametrical Lie group. We will see that, in addition, it admits a hidden supersymmetry.

In view of the rotational invariance of equation (\ref{equ5}) and
its symmetry with respect to shifts of time variable, it is possible
to search for its solution in separated variables. In particular, we
can use the spherical variables and search for solutions in the
following form:
\begin{gather}\la{psi}\Psi=\text{e}^{-iEt}R_{lm}( r)Y_{lm}(\theta,\varphi)\end{gather}
where $\theta$ and $\varphi$ are angular variables and
$Y_{lm}(\theta,\varphi)$ are spherical functions, i.e., eigenvectors
of $L^2=L_1^2+L_2^2+L_3^2$ and $L_3$. As a result we come to the
following radial equation for $R_{lm}$:
\begin{gather}\la{equ7}-r^{2\nu+2}\frac{\p^2R_{lm}}{\p r^2}-(2\nu+4)r^{2\nu+1}\frac{\p R_{lm}}{\p r} +
\left(r^{2\nu}({l(l+1)+\kappa})+\omega^2r^{-2\nu}\right)R_{lm}
=2ER_{lm}.\end{gather}

This equation can be simplified using the Liouville transform:
\begin{gather}\la{equ8}r\to z=r^{-\nu},\quad R_{lm}\to\tilde R_{lm}=z^{\frac{\nu+3}{2\nu}}R_{lm}\end{gather} which reduce (\ref{equ7}) to the following form:
 \begin{gather}\la{equ9}-\nu^2\frac{\p^2\tilde R_{lm}}{\p z^2}+\left(\frac{l(l+1)+\delta}{z^2}+\omega^2z^2\right)\tilde R_{lm}
=2E\tilde R_{lm}\end{gather}
where we denote
\begin{gather*} \delta=\frac3{4}(\nu+1)(\nu+3)+
2\kappa.\end{gather*}

Equation (\ref{equ9}) describes a deformed 3d harmonic oscillator which includes two deformation parameters $\nu$ and $\kappa$. To justify such interpretation consider first the special case when these parameters are constrained by the following condition:
\begin{gather}\la{con}2\kappa=-\nu^2-3\nu-2,\end{gather}
and equation (\ref{equ9}) is reduced to the following form:
 \begin{gather}\la{eg1}H_l\tilde R_{lm}\equiv\left(-\nu^2\frac{\p^2}{\p z^2}+\frac{(2l+1)^2
 {-\nu^2}}{4z^2}+\omega^2z^2\right)\tilde R_{lm}
=2E\tilde R_{lm}.\end{gather}

Equation (\ref{eg1}) is shape invariant. Indeed, the Hamiltonian
$H_r$ can be factorized:
\begin{gather}H_l=a_l^+a_l-C_l\la{eg2}\end{gather}
where
\begin{gather}\la{eg3}\begin{split}&a=-\nu\frac{\p}{\p z}+W,\quad a^+=\nu\frac{\p}{\p z}+W,\\&W=\frac{2l+1+\nu}{2z}+\omega z,\quad C_l=\omega(2l+2\nu+1).\end{split}\end{gather}

On the other hand, the superpartner $\hat H_l $ of Hamiltonian
(\ref{eg2}) has the following property which fixes the shape
invariance:
\begin{gather}\hat H_l\equiv a_la_l^++C_l=H_{l+\nu}+C_l.\la{eg4}\end{gather}

 Shape invariant equations can be solved algebraically using tools of SUSY quantum mechanics, see, e.g., \cite{khare}. In particular, for equation (\ref{eg1}) the ground state energy level $E=E_0$ is equal to $\frac12C_l$, while  for the n-th exited state we obtain $E_n=\frac12C_{l+\nu n}$, or
 \begin{gather}\la{eg6}E_n=\omega\left(2n\nu+l+
 \nu+\frac12\right)=\omega\left(2n+l+\frac32\right)+\delta\omega(2n+1)\end{gather}
 where $\delta=\nu-1$.

 Equation (\ref{eg6}) represents the spectrum of 3d isotropic harmonic oscillator deformed by the term proportional to $\delta$. For integer $\delta$ the spectrum is degenerated, but this degeneracy is much less than in the case of the standard 3d oscillator which corresponds to $\delta =0$.

Equation (\ref{equ9}) is shape invariant also in the general case
when condition (\ref{con}) is not imposed. The corresponding energy spectrum is:
\begin{gather}\la{spect}E_n=\frac\omega2(\nu(2n+1)+
\sqrt{(2l+1)^2+\tilde \kappa}\ )\end{gather}
where $\tilde \kappa=8(\kappa+1)+\nu(\nu+3).$ The related eigenvectors are expressed via the confluent hypergeometric functions $\cal F$:
\begin{gather}\la{evec} R_n=\text{e}^{-\frac{\omega r^{\nu}}{2\nu}}r^{\nu n-\frac{E_n}{\omega}}{\cal F}\left(-n, \frac{E_n}{\nu\omega}-n,\frac{\omega}{\nu} r^{-\nu}\right)\end{gather}
where $n$ is integer and $E_n$ is the corresponding eigenvalue given
by equation  (\ref{spect}).

Thus we effectively solve equation (\ref{equ5}) using tools of SUSY
quantum mechanics. However, there is one more way to do it, as it
will be shown in the next section.

\subsection{Shape invariance with Morse potential}
Let us return to equation (\ref{equ7}) and solve it using more sophisticated approach. First we divide all terms by $r^{2\nu}$ and obtain:
\begin{gather}\la{equq1}-r^{2}\frac{\p^2R_{lm}}{\p r^2}-(2\nu+4)r\frac{\p R_{lm}}{\p r} +
\left(\omega^2r^{-4\nu}
+\mu r^{-2\nu}\right)R_{lm}
=\varepsilon R_{lm}\end{gather}
where
\begin{gather}
\la{B}\varepsilon={-l(l+1)-2\kappa}, \quad \mu=-2E.\end{gather}

In equation (\ref{equq1}) we change the roles of the coupling
constant and energy and consider it as an eigenvalue problem for
$\varepsilon$. Making the Liouville transform
\begin{gather}\la{equq2}r\to \rho=\ln(r),\quad R_{lm}\to\tilde R_{lm}=\text{e}^{-\frac{\nu+3}2}R_{lm}\end{gather}
we simplify (\ref{equq1}) to the following form:
\begin{gather}\la{equq3}H_\nu\tilde R_{lm}\equiv\left(-\frac{\p^2}{\p \rho^2}+\omega^2\text{e}^{-2\nu\rho}+
(\omega\nu+\omega{\nu})\text{e}^{-\nu\rho}\right)
\tilde R_{lm}=\hat\varepsilon\tilde R_{lm}\end{gather}
where
\begin{gather}\la{A}\hat\varepsilon=\varepsilon-
\left(\frac{\nu+3}2\right)^2 ,\quad \nu=\frac{\mu}{\omega}-\frac{\nu}2\end{gather}

Equation (\ref{equq3}) includes the familiar Morse potential and so is shape invariant. Indeed, denoting $\mu=\omega(\nu+\frac\nu2)$  we can factorize hamiltonian $H_\nu$ like it was done in (\ref{eg2}) where index $l$ should be changed to $\nu$ and
\begin{gather}\la{w}W=\nu-\omega\text{e}^{-a\rho},\quad C_\nu=\nu^2\end{gather}
and the shape invariance is easy recognized.

To find the admissible  eigenvalues $\varepsilon$ and the corresponding
eigenvectors we can directly use the results presented in paper \cite{khare}, see Item 4 of
Table 4.1 there:
\begin{gather}\la{C}\begin{split}& \hat\varepsilon =\hat\varepsilon_n=-(\nu-n\nu)^2,\\&
(\tilde R_{lm})_n=y^{\frac\nu\nu-
n}\text{e}^{-\frac{y}2}L_n^{2(\frac\nu\nu-n)}
(y)\end{split}\end{gather}
where $y=\frac{\omega}\nu r^{-\nu}$.

Thus we find the admissible values of $\hat\varepsilon_n$. Using
definitions (\ref{B}) and (\ref{A}) we can find the corresponding
values of $E$ which are in perfect accordance with (\ref{spect}).
\section{Discussion}
It is an element of common knowledge that PDM Schr\"odinger
equations are not invariant with respect to Galilei transformations,
see discussion of this point in \cite{LL}. A natural question is:
what do we have instead, what kinds of symmetry do exist in PDM
systems?

The results presented above in Sections 6 and 7 give a constructive
answer to this question. They include the completed list of
continuous symmetries which can be admitted by PDM Schr\"odinger
equations. In particular, the systems presented in Items 1-4 of
Table 5 admit Galilei transformation of variable $x_3$, provided the
time variable is changed also. Indeed, integrating the Lie equations
corresponding to generators $Q^3_1$, we obtain the following
symmetry transformations for the independent variables:
\begin{gather}\la{st}x_3\to x_3'= x_3+vt, \quad t\to t'=\frac{t}{1-vt},\quad x_1\to x_1'=x_1,
\quad \quad x_2\to x_2'=x_2.\end{gather}

Solving the Lie equations, it is not difficult to find also the
corresponding transformation low for the wave function $\psi$, but
we omit the related cumbersome formula.

Thus we extend the well known results \cite{Boy} to the case of
Schr\"odinger equations with position dependent mass. The number of
PDM systems with non-equivalent symmetries appears to be much more
large than in the case of constant masses. However, the maximal
admissible symmetries  are less extended. In particular, we cannot
find PDM systems invariant w.r.t. ten parametrical Galilei group or
twelve parametrical extended Galilei group which are admissible by
the constant mass Schr\"odinger equation with harmonic oscillator,
linear and constant potentials. Nevertheless, there are   PDM
systems admitting eight, seven or six parameter symmetry groups, see
Section 7 for discussion of this point. In addition, thanks to the
linearity of the considered equations, there is the infinite
symmetry group of transformations (\ref{ququ}).

As  other extensions of results of paper \cite{Boy} we can mention
the group classification of the nonlinear Schr\"odinger equations \cite{Pop}
 and the analysis of its conditional symmetries \cite{FN}.

To find the determining equations for symmetries of equations (\ref{A1}),
(\ref{A11}) we use the traditional technics applied in papers
\cite{Hag}--\cite{Boy}, i.e., evaluate commutators of the searched symmetry
operators with $\ri\p_t-H$, see Section 3. The more general approach which
is applicable also to nonlinear equations was developed long time ago by
Sophus Lie, see, e.g., the fundamental Olver monograph \cite{olver}.
But we choose a more simple way which does not request a knowledge of  the
group analysis grounds from potential readers of the present paper. Let us
note that this way is applicable also for group analysis of some non-linear
equations \cite{N6}, \cite{N1}.

However, the deducing of the determining equations is a necessary
but rather simple step in the group classification of a given class
of equations.  Much more efforts are requested to obtain the
complete set of inequivalent solutions of the determining equations.
To achieve this goal we apply the algebraic approach  which includes
the description of  subgroups of the equivalence group  of equations
(\ref{seq}) and a priory analysis of the admissible invariance
groups. The grounds of the algebraic approach were created in papers
\cite{Renat}, but in fact its elements  were used for the first time
in paper \cite{gan}.

To obtain the completed group classification of equations
(\ref{seq}) it was necessary to describe also the equivalence
transformations which keep the generic form of these equations but
can change the explicit form of arbitrary functions $f$ and $V$. It
is well known that in general the equivalence transformations do not
form a Lie group, but have a structure of groupoid \cite{popa}. This
is the case also for equations (\ref{seq}), and for some subclasses
of these equations  there exist   additional equivalence relations
which do not belong to their equivalence group.  These relations are
specified in the tables heads and can be used to transform some of
the presented systems to more simple ones, like the isotropic
harmonic oscillator can be reduced to the Schr\"odinger equation
with the trivial potential.

A special attention was paid to the equivalence relations of the
ambiguity parameters. In Section 8 we present the notion of the
ambiguity invariance of PDM Hamiltonians which is defined as the
invariance w.r.t. changes of the ambiguity parameters up to a
constant potential term. In this section extended classes of
Hamiltonians having this property are specified.

Thanks to their extended symmetries many of the presented systems
are exactly solvable. In Section 10 an example of a solvable system
is discussed. In addition to the symmetry under the six parameter
Lie group whose generators are given in Item 18 of Table 3 and
formula (\ref{P0}), equation  (\ref{equ5}) possesses a hidden
dynamical symmetry w.r.t. group SO(1,2). The effective radial
hamiltonian is shape invariant, and its eigenvalues can be easily
found with tools of SUSY quantum mechanics. We call this system
deformed 3d isotropic harmonic oscillator. In spite on the
qualitative difference of its spectra  (\ref{eg6}) and (\ref{spect})
of the standard 3d oscillator, it keeps the main supersymmetric
properties of the latter.

We discuss SUSY aspects only of one selected system. But in fact the
classification tables present a  number of supersymmetric systems with
hidden dynamical symmetries. In particular, there are other systems with
supersymmetric radial equations, systems with supersymmetric  equations in
angular variables and systems with supersymmetries of both the mentioned
types. To keep the reasonable paper size we do not discuss SUSY aspects of
systems other than equation (\ref{eg6}).

The very possibility to solve shape invariant Schr\"odinger equation
with using more than one superpotential is well known, see, e.g.,
\cite{Co}. We show that it is possible to do it with the PDM system
(\ref{eg6}) by changing the roles of the energy eigenvalue and
coupling constant.

In the present paper we did not consider  Schr\"odinger equations
with constant mass. However, such equations can appear as particular
case of systems presented in the classification tables.  Examining
these cases  we found some systems missing in Boyer classification
\cite{Boy}, refer to Section 9 in the above.

It is necessary to stress that symmetries of the constant mass Schr\"odinger
 equation cannot be completely described in frames of the more general
 problem of group classification of PDM Schr\"odinger equations, since the
 latter includes  much more extended equivalence group which reduces the
 number of non-equivalent symmetries. Thus the revision of classical
 results presented in \cite{Boy} is requested, but it deserves a separate
 publication which is in preparation.

 In the present paper the continuous  symmetries of PDM Schr\"odinger
 equations are discussed. We do not consider {\it discrete}
 symmetries which present additional powerful tools in construction
 of supersymmetric and exactly solvable  models with constant masses \cite{AN1},
 \cite{AN2}. Moreover, the discrete symmetry operators can serve as
 constructive elements of physically consistent potentials \cite{vin1}, \cite{vin2}.
 A systematic study of discrete  symmetries of PDM Schr\"odinger
 equations is one
 more interesting  field which still
 waits for researchers attention.

\renewcommand{\theequation}{A\arabic{equation}} %
\setcounter{equation}{0}
\appendix
\section{Appendix}
\subsection{ Time dependence of symmetry operators}
Here we analyze the compatibility condition (\ref{5.8}) and prove
that the time dependence of symmetry operators with non-trivial
$\dot\xi^a$ is described by  formulae (\ref{e71})-- (\ref{e73}) and
(\ref{q7})--(\ref{q9}).

Let us differentiate (\ref{kil}) w.r.t. $t$ and substitute the
obtained expression for $\dot\xi^a$ into (\ref{5.8}). Then, equating
the coefficients for the same powers of $x^a$ we obtain the
following conditions for functions $\dot\lambda^a,\ \dot\nu^a,\
\dot\theta^a,\ a$ and $\dot \omega$:
\begin{gather}\la{10}\dot\lambda^a=\dot\nu^a=0,\quad \dot \omega=0, \quad a,b=1,2,3.\end{gather}
or, alternatively,
\begin{gather}\la{5.9}\dot \theta^a=0,\quad \dot\lambda^a\dot\nu^b-
\dot\lambda^b\dot\nu^a=0,\quad a=1,2,3\end{gather}

 Considering the case (\ref{10}) we come to the following form of the doted symmetry operator:
\begin{gather}\la{5.17}\dot Q=\dot\xi^0\p_0+\varepsilon^{abc}\dot\theta^ax^b\p_c,\end{gather}

Let conditions (\ref{5.9}) are valid then functions $\dot\lambda^a$
and $\dot\nu^a$ can be represented in the following form
\begin{gather}\dot\lambda^a=g\tilde\mu^a,\quad \dot \nu^a=q\tilde\mu^a\la{11}\end{gather}
where $g,\ q$ and $\tilde\mu^a$ are some functions of $t$. In
accordance with (\ref{kil}), (\ref{5.9}) and (\ref{11}) the
corresponding vector $\dot\xi^a$ is reduced to:
\begin{gather}\la{12}\dot\xi^a=\tilde\mu^a(gr^2+b)+x^a(\dot \omega-2g\tilde\mu^bx^b)\end{gather} and can be correlated to the following symmetry operator (\ref{so}) differentiated with respect to time:
\begin{gather}\begin{split}&\la{5.13}\dot Q=\ri\dot\xi^0\p_0+\tilde\mu^a
(gK^a+qP_a)+\dot\omega
D+\dot\eta\\&\equiv\ri\dot\xi^0\p_0+\tilde\mu^a\left((g+q)M_{0a}+
(g-q)M_{4a}\right)+\dot\omega M_{04}+\dot
\eta\end{split}\end{gather} where relations (\ref{iso}) were used.

Making hyperbolic rotations on the plane 0--4, it is possible to
reduce (\ref{5.13}) to one of the following forms:
\begin{gather}\la{5.16}
\dot Q=\ri\dot\xi^0\p_0+\mu^a(M_{0a}-M_{4a}) +\dot\omega
M_{04},\quad \text {if}\quad (g+q)^2=(g-q)^2,\\\la{5.14}\dot
Q=\ri\dot\xi^0\p_0+\mu^aM_{0a}+ \dot\omega M_{04},\quad \
\phantom{aaaaaaa}\text {if}\quad (g+q)^2>(g-q)^2,\\\dot
Q=\ri\dot\xi^0\p_0+\mu^aM_{4a} +\dot\omega M_{04},\quad \
\phantom{aaaaaaa}\text {if}\quad (g+q)^2<(g-q)^2
\la{5.15}\end{gather} where $ \mu^a=2\tilde\mu^a (g^2+q^2)^\frac12$.

Thus our analysis of time dependent symmetries is reduced to
considering the four independent cases presented by formulae
(\ref{5.17}), (\ref{5.16}),  (\ref{5.14}) and (\ref{5.15}).

Let us start with symmetries defined by relation (\ref{5.17}). The
corresponding vector $\dot\xi^a$ is:
\begin{gather}\la{e2}\dot\xi^a=\varepsilon^{abc}x^b\dot\theta^c\end{gather}
which leads to the following  equations (\ref{5.4}) and (\ref{5.6}):
\begin{gather}\begin{split}&\dot\xi^a\hat f_a=\dot a,
\\&\dot\xi^a\hat f_b-\dot\xi^b\hat f_a=-2\varepsilon^{abc}\dot\theta^c.
\la{e3}\end{split}\end{gather}

Solving (\ref{e3}) for $\hat f_a$ we obtain:
\begin{gather}\la{e4}\hat f_a=\frac{2(x^a\dot\theta^2-\dot\theta^ax^b
\dot\theta^b)+ \dot
a\varepsilon^{abc}x^b\dot\theta^c}{x^bx^b\dot\theta^2-
(x^b\dot\theta^b)^2}.\end{gather}

Two evident consequences of equation (\ref{e4}) are:
\begin{gather}\la{e5}\dot\theta^b\hat f_b=0, \quad x^b\hat f_b=2.\end{gather}

In order to systems of equations (\ref{e4}) and (\ref{e5}) be
compatible, all  functions $\dot\theta^1, \ \dot\theta^2,\
\dot\theta^3$ and $\dot a$ have to be linearly dependent, i.e.
proportional to the same function of $t$.
  Thus, up to rotation
transformations, we can restrict ourselves to the case when
$\dot\theta^1=\dot\theta^2=0,\ \dot a=\bar a\theta^3 $ and reduce
equations (\ref{e3}) to the following form:
 \begin{gather}\la{e7}\dot \xi^1=-\theta x^2,\quad \dot
\xi^2=\theta x^1,\quad \dot \xi^3=0\end{gather} where
$\theta=\left((\theta^1)^2+(\theta^2)^2+(\theta^3)^2\right)^{\frac12}$
is a function of $t$.

 Thus the version (\ref{5.17}) can be effectively reduced to the case
 when vector $\dot\xi^a$ is given by equations (\ref{e71}) where we denote
 $\theta=\dot\Phi$.

 Let us consider the next version of doted symmetry operators given by equation (\ref{5.17}). The corresponding vector $\dot\xi^a$ looks as follows
\begin{gather}\la{eq4}\dot \xi^a =\mu^a+\dot\omega x^a\end{gather}
and so equations (\ref{5.4}) and (\ref{5.6}) are reduced to the
following system:
\begin{gather}\la{eq1}\dot\xi^a\hat f_a=\dot a+2\dot\omega,\\\la{eq2}\dot\xi^a\hat f_b-\dot\xi^b\hat f_a=0.\end{gather}

Multiplying (\ref{eq2}) by $\dot\xi^b$ and summing up w.r.t. index
$b$ we obtain:
\begin{gather}\la{eq3}\hat f_a{\dot\xi^b\dot\xi^b}=\dot \xi^a(\dot a+
2\dot\omega).\end{gather}

Differentiating equation (\ref{eq3}) w.r.t. $t$ and taking into
account time independence of $\hat f_a$, we obtain the following
differential consequence:
\begin{gather}\la{eq5} {\dot\xi^b\dot\xi^b} ({\ddot
\xi^a(\dot a+2\dot\omega)}+\dot \xi^a(\ddot a+2\ddot\omega))= 2{\dot
\xi^a(\dot a+2\dot\omega)}\ddot\xi^b\dot\xi^b.\end{gather}

Substituting (\ref{eq4}) into (\ref{eq5}) and equating the
coefficients for the same powers of independent variables $x^a$ we
come to the following system of equations for functions $\mu^a$,
$\dot\omega$ and $\tau=\dot a+2\dot \omega$:
\begin{gather}\la{eq6}\tau{\ddot \omega}={\dot \omega}{\dot \tau}; \ \ \ \dot\mu^a\tau=\mu^a\dot\tau.\end{gather}

In accordance with (\ref{eq6}) functions $\mu^a,\ \dot\omega$ and
$\tau$ should either be zero or proportional one to another. Since
zero value of $\tau$ corresponds to constant masses (see equation
(\ref{eq3})), we suppose $\tau\neq0$. Then  equations (\ref{eq6}) present the following alternatives:
\begin{gather}\la{eq7}\dot\omega=0,\quad \mu^a=aC_a;
\\\la{eq8}\dot\omega=C_0,\quad \mu^a=C_a;\\\la{eq9}\ddot\omega\neq0,
\quad \tau=C_0\dot\omega,\quad \mu^a=C_a\dot\omega\end{gather} where
$C_0$ and $C_a$, $a=1,2,3$ are constants.

For both versions  (\ref{eq8}) and (\ref{eq9}) the corresponding
functions (\ref{eq4}) can be reduced to the  form given in (\ref{e72}).
To this effect it
is sufficient to make the shifts $x_a\to
x_a-\frac{C_a}{\dot\omega}$.

If version (\ref{eq7}) be realized, then up to rotation
transformations we can restrict ourselves to functions $\dot \xi^a$
given by  formula (\ref{e73}).

Consider now the remaining versions (\ref{5.14}) and (\ref{5.15}).
In both cases the corresponding vectors $\dot\xi^a$ (\ref{12}) can
be represented in the following unified form:
\begin{gather}\la{e8}\dot\xi^a=\mu^a(r^2+\delta)-2x_a\mu^bx_b+\dot\omega x_a\end{gather}
were $\delta=\pm1$. The corresponding equations (\ref{5.4}) and
(\ref{5.6}) are reduced to the following system:
\begin{gather}\la{e9}\begin{split}&\dot\xi^a\hat
f_a= \dot a+2\dot\omega-4\mu^ax_a,\\&\dot\xi^a\hat f^b-\dot\xi^b\hat
f_a=4(\lambda^ax_b-\lambda^bx_a).\end{split}\end{gather} Solving
these equations for $\hat f_a$ we obtain:
\begin{gather}\hat f_a=\frac{\mu^aR+x_aP}{\dot \xi^b\dot\xi^b}\la{6}\end{gather}
where
\begin{gather}\begin{split}&\la{161}R=r^2(\dot a-2\dot \omega)-8\delta\mu^cx_c+\delta(\dot a+2\dot\omega),\\&P=(\dot\omega- \mu^cx_c)(\dot a+2\dot\omega)+4\mu^c\mu^c(r^2+\delta),\\&\dot \xi^b\dot\xi^b=\mu^c\mu^c(r^2+\delta)^2-4(\delta+\dot \omega(r^2+\delta))(\mu^cx_c)^2+2\dot\omega(\delta-r^2)\mu^cx_c+ \dot \omega^2r^2\end{split}\end{gather}

 Differentiating (\ref{6}) with respect to $t$ and and taking into account time independence of $\hat f_a$, we obtain the following condition:
 \begin{gather}\la{17}(\dot\mu^aR+\mu^a\dot R+x_a\dot P)\dot\xi^b\dot\xi^b=2(\mu^aR+x_aP)\ddot\xi^b\dot\xi^b.\end{gather}
Substituting expressions (\ref{161}) into (\ref{17}) and equating the
coefficients for the same powers of $x_a$ we conclude that all
functions $\mu^a$ and $\dot \omega$ should be proportional to the
same function of $t$ which we denote as $\ddot \Phi$, i.e.,
\begin{gather}\la{18}\mu^a=C_a\ddot\Phi(t),\quad \dot\omega=C_0\ddot\Phi(t).\end{gather}
The corresponding operators (\ref{5.14}) and (\ref{5.15})   are:
\begin{gather}\la{19}\dot Q=\dot\xi^0\p_0+\ddot\Phi(t)(C_aM_{0a}+
C_0 M_{04})\end{gather} and
  \begin{gather}\la{20}\dot Q=\dot\xi^0\p_0+\ddot\Phi(t)(C_aM_{4a}
+C_0 M_{04}).\end{gather}

Using the rotation transformations we can reduce nontrivial linear
combinations $C_aM_{0a}$ of vectors $M_{0a}$ to $CM_{03}$ with
$C^2=C_1^2+C_2^2+C_3^2$. Then making a planar rotation on the plane
3--4, it is possible to reduce the linear combination $C_0
M_{04}+CM_{03}$ to $\tilde CM_{04}$ with $\tilde C^2=C^2+C_0^2$. As
a result we reduce (\ref{19}) to the case with trivial $C_a$, and
the corresponding vector $\dot\xi^a$ again takes the form given by
equation (\ref{e72}).

Analogous speculations can be applied to operator (\ref{20}) which
can be reduced to the case of trivial $C_a$ provided $C_0^2>\tilde
C^2$. If $C_0^2<\tilde C_2$ we can reduce this operator to the form
(\ref{5.17}), while for $C_0^2=\tilde C^2$ it is reducible to the
particular case  (\ref{5.16}) with $\dot\omega=0$.

Thus we prove that the classifications of symmetries with time
dependent Killing vectors $\xi^a$ can be reduced to solving the
determining equations (\ref{5.1})--(\ref{5.3}) with $\dot\xi^a$
enumerated in formulae (\ref{e71}), (\ref{e72}) and (\ref{e73}).
\subsection{Symmetries with time dependent derivative terms} Let us start with
symmetries of the first class when functions $\xi^a$ are time
dependent and have one of the forms presented in
(\ref{e71})--(\ref{e73}).

For functions (\ref{e71}) the determining equations
(\ref{5.1})--(\ref{5.3}) are reduced to the following form:
\begin{gather}\la{q1}\eta_{1}=-\frac{\dot \Phi x_2}{2f},\quad
\eta_{2}=\frac{\dot \Phi x_1}{2f},\quad
\eta_3=0,\\\la{q2}\Phi(x_1\hat f_2-x_2\hat
f_1)=a,\\\la{q3}\Phi(x_1V_2-x_2V_1)=a V+\dot\eta.
\end{gather}

The generic solution of equation (\ref{q2}) looks as follows:
\begin{gather*}f=e^{\nu\Theta}F(\tilde r),\quad a=\nu\Phi\end{gather*}
where $\tilde r=\sqrt{x_1^2+x_2^2}, \
\Theta=\arctan\left(\frac{x_2}{x_1}\right)$, $\nu$ is a constant,
and $F$ is an arbitrary function of $\tilde r$. This solution is
compatible with equation (\ref{q1}) iff $F=\mu\tilde r^2$ with some
constant $\mu$, which can be reduced to the unity by rescaling the
time variable. In this way we obtain
\begin{gather}\la{q4} f=\tilde r^2e^{\nu \Theta}, \quad \eta=
\frac1{\nu}\dot\Phi e^{\nu\Theta}+\phi(t),\quad \nu\neq0.
\end{gather}

Substituting (\ref{q4}) into (\ref{q3}) we obtain:
\begin{gather}\la{q5}\Phi(x_1V_2-x_2V_1-\nu V)=\frac1{\nu}
\ddot \Phi e^{-\nu\Theta}+\dot\phi(t).\end{gather}

This equation specifies admissible functions $\Phi$ and $\phi$.
Indeed, to make  it  consistent we have to set
\begin{gather}\la{coco}\ddot\Phi=\kappa\Phi, \quad
\phi=\nu\Phi\end{gather} where $\kappa$ and $\nu$ are constants.
Then substituting (\ref{coco}) into (\ref{q5}) we obtain the
following equations for potential $V$:
\begin{gather}\la{q6}x_1V_2-x_2V_1-\nu V=\frac\kappa{\nu}
e^{-\nu\Theta}+\nu.\end{gather}

Since $\nu$ is nonzero, it is possible it is possible to reduce
$\nu$ to zero by  a constant shift of the potential.  In addition,
the generic solutions of the first of equations (\ref{coco}) is
given by formulae (\ref{q7})--(\ref{q9}).

Formulae (\ref{q4}), (\ref{e71}), (\ref{q7})--(\ref{q9}) present
explicitly all components of symmetry operators (\ref{so}) together
with the corresponding inverse mass function $f$. Solving equation
(\ref{q6}) consequently for zero, negative and positive $\kappa$ we
obtain the corresponding potentials:
\begin{gather}\la{pot1}V=G(\tilde r,x_3)e^{\sigma \Theta}
\phantom{aaaaaaaaaaaaaaaaaa} \texttt{if}\quad
\kappa=0,\\\la{pot2}V=G(\tilde
r,x_3)e^{\sigma\Theta}+\frac{\lambda^2}{2\mu\sigma^2}
e^{-\sigma\Theta}, \quad \texttt{if}\quad
\kappa=-\lambda^2,\\\la{pot3}V=G(\tilde
r,x_3)e^{\sigma\Theta}-\frac{\lambda^2}{2\mu\sigma^2}
e^{-\sigma\Theta}, \quad \texttt{if}\quad
\kappa=\lambda^2.\end{gather}

In the special case $\sigma=0$ relations (\ref{q4}) are transformed
to the following form:
\begin{gather}\la{q10}\Phi(x_1V_2-x_2V_1)=\frac12\ddot\Phi\Theta+\dot\phi,\\\la{q11}f=\tilde r^2,\quad
\eta=\frac12\dot\Phi\Theta+\phi(t).\end{gather}

Functions $\Phi$ and $\phi$ should satisfy relations (\ref{coco}) ,
and so equation (\ref{q10}) is solved by the following function:
\begin{gather}\la{q12}V=\kappa
\Theta^2+\sigma\Theta+G(\tilde r,x_3).\end{gather}

Thus we have specified all Hamiltonians which admits symmetries of
special form (\ref{so}), (\ref{e71}). Any of such Hamiltonians
admits  the following representation:
\begin{gather}H=p_a\tilde
r^2e^{\sigma\Theta}p_a+V\la{f5}\end{gather} where arbitrary
parameter $\sigma$ can take both the zero and nonzero values and
possible  potentials $V$ are given by formulae (\ref{pot1}),
(\ref{pot2}), (\ref{pot3}) and (\ref{q12}).
 The corresponding functions $\xi^0$ present in symmetry
operator (\ref{so}) are easily calculated solving  equation
(\ref{de1}).

Such specified Hamiltonians include arbitrary functions $G(\tilde
r,x_3)$.  It is important to note that for some fixed $G(\tilde
r,x_3)$ the corresponding equation (\ref{seq}), (\ref{f5}) can have
additional symmetries. To complete the group classification of the
considered subclass of equations it is necessary to specify
Hamiltonians (\ref{f5})  compatible with  all possible nonequivalent
additional symmetries.  This problem is rather simple since the set
of transformations which keep the fixed form of $f$ and the generic
forms of potentials (\ref{pot2}), (\ref{pot3}) and (\ref{q12}) is
rather restricted. Namely, we can make only shifts of $x_3$,
simultaneous scalings of all spatial variables, and combinations of
the conformal and shift transformations generated by operator
$M_{43}$ (we call them "conformal+shift transformations"). Moreover,
there are five inequivalent possibilities: only shifts, only
scalings, only conformal+shift transformations, and all
transformations mentioned in the above. In order to these
transformations be admissible,  functions $G(\tilde r,x_3)$ in
(\ref{pot1}), (\ref{pot2}), (\ref{pot3}) and (\ref{q12}) have to be
invariant with respect to these transformations, i.e., there are
four versions: $G=G(\tilde r)$ for equations invariant w.r.t. shifts
of $x_3$, $G=G(\frac{x_3}{ r})$ for scalings, $G=G(\frac{r^2+1}{
\tilde r})$ for transformations generated by $M_{43}$, and $G=Const$
if all mentioned transformations  are acceptable.  The corresponding
versions of  inverse masses and potentials together with the related
symmetries  are presented in Tables 1 and 2.

In complete analogy with the above we can solve determining
equations (\ref{5.1})--(\ref{5.3}) for special cases of functions
$\xi^2$ presented in (\ref{e72}) and (\ref{e73}). The corresponding
classification results are presented in Tables 3 and
4.
\subsection{Symmetries of classes 2 and 3}
For symmetries (\ref{so}) with time independent coefficients $\xi^a$ we can
directly use the results of paper \cite{Pate} concerning the subgroup structure of group SO(1,4) and decouple the system of determining equations  (\ref{5.1})--(\ref{5.3}) to inequivalent subsystems corresponding to selected values of arbitrary parameters in vector (\ref{kil}), like it was done in \cite{NZ}. The only new feature in comparison with \cite{NZ} is the necessity to take into account additional time dependent  functions $\xi^0$ and $a$ which appear now in the determining equations.

Let us start with one dimensional subalgebras of algebra c(3)$\thicksim$so(1,4). In accordance with \cite{Pate} it is possible to specify five inequivalent subalgebras of the mentioned type spanned on the following basis elements:
\begin{gather}\la{be}\begin{split}&  \langle P_3\rangle,\quad \langle L_3 \rangle, \quad \langle P_3+L_3\rangle,\\&\langle K_3-P_3+\nu L_3\rangle,\quad 0\leq\nu\leq1,\quad \langle \kappa L_3 +D\rangle,\ \ 0<\kappa\leq 1.\end{split}\end{gather}

Subalgebra spanned on $P_3$ is associated with the following symmetry operator
\begin{gather*}Q=\xi^0+\partial_3+\ri\eta\end{gather*}
where $\xi^0$ and $\eta$ are functions of $t$ and $t, \bf x$ respectively.
The corresponding determining equations (\ref{de1}) and (\ref{5.1})--(\ref{5.3}) are reduced to the following system:
\begin{gather}\la{detete}\begin{split}&f_3=a f,\ \ a=-\dot\xi^0,\\&\eta_a=0, \ \ V_3=aV+\dot \eta.\end{split}\end{gather}

Taking into account time independence of $f$ and $V$, we conclude that $\xi^0$ and $\eta$ are linear functions of $t$, and $a$ is a constant:
\begin{gather*}\xi^0=\nu t+\mu,\ \ \eta=\kappa t +\rho, \quad a=-\nu\la{opa}\end{gather*}

 We can restrict ourselves to the case $\rho=\mu=0$ since these parameters are coefficients for symmetries accepted by any equation (\ref{seq}). In addition, if $\sigma$ is nontrivial, we can set   $\kappa=0$ since this parameter can be removed transforming $Q\to\text{e}^{-\ri\frac{\kappa}{\sigma}t}Q\text{e}^{\ri\frac{\kappa}{\sigma}t}.$
 Thus effectively there are only two versions:
 \begin{gather}\la{ver1}\xi^0=\nu t,\ \eta=0\quad \text{ and } \quad
 \xi^0=0,\ \eta=\kappa t.\end{gather}
  Solutions of the corresponding equations (\ref{detete}) are:
 \begin{gather}\la{sos1}f=F(x_1,x_2)e^{\nu x_3},\
 V=G(x_1,x_2)e^{\nu x_3}\quad \text{ if }\quad Q=P_3+\ri\nu t\p_t \end{gather}
 and
 \begin{gather}\la{sos2}f=F(x_1,x_2),\ V=G(x_1,x_2)+\kappa x_3\quad \text{ if }\quad Q=P_3+\kappa t.\end{gather}
 Just these solutions are
 represented in Items 1 and 2 of Table 6.

 In complete analogy with the above we can solve the determining equations corresponding to the other one dimension algebras presented in (\ref{be}).

Let us consider two-dimensional subalgebras. In accordance with \cite{Pate} it is sufficient to specify five of them:
\begin{gather}\la{beba}\begin{split}& \langle P_3, P_1\rangle,\quad  \langle P_3, L_3 \rangle,\quad  \langle P_3, D+\kappa L_3 \rangle,\quad
\langle D, L_3\rangle, \quad
\langle L_3, K_3-P_3\rangle, .\end{split}\end{gather}

All algebras whose basis elements are presented in (\ref{beba}) include one dimensional subalgebras fixed in (\ref{be}). Functions $f$ and $V$ which correspond to the one dimensional subalgebras  are known and presented in Items 1--10 of Table 6. Let $f_{(1)}$ and $V_{(1)}$ correspond to the first  elements of pairs given in (\ref{beba}).
   Our task is to  substitute these functions   into the determining equations  generated by the second elements and integrate the obtained system.

   For algebras presented in the first line of equation (\ref{beba}) we have functions  $f=f_{(1)}$ and $V=V_{(1)}$ given by equations (\ref{sos1}) and (\ref{sos2}). The determining equations corresponding to the second basis element $Q=P_2$ can be obtained from (\ref{detete}) by changing the subindex 3 to 1.

   We have again two possibilities fixed in (\ref{ver1}), so the second symmetry operator can be reduced to one of the following forms:
   \begin{gather}\la{sss}\tilde Q=P_1+\ri\mu t\p_t, \quad \text{ or }\quad \tilde Q=P_1+\nu t.\end{gather}

   There are four different pairs $\langle Q,\ \tilde Q\rangle $ with $Q$ and $\tilde Q$ given in equations (\ref{sos1}), (\ref{sos2}) and (\ref{sss}). Moreover, passing to a new basis in algebras $\langle Q,\ \tilde Q\rangle $ and making the suitable linear transformations of variables $x_1$ and $x_2$ it is possible to reduce them to the following two representatives:   $\langle P_3+\kappa t,\ P_1+\ri\sigma t\p_t\rangle $ and $\langle P_3+\ri\sigma t\p_t,\  P_1+\kappa t\rangle.$
   functions $f$ and $V$ have to solve the following systems of equations: \begin{gather}f_1=\nu f,\quad V_1=\nu V\la{su1}\end{gather}
   for functions (\ref{sos2}), and
   \begin{gather}f_1=0,\quad V_1=\mu \la{su}\end{gather}
for functions (\ref{sos1}). As a result we obtain solutions
represented in Items 15 and 16 of Table 6.

Consider now the second algebra from the list (\ref{eg6}). Its basis element $L_3$ is associated with the following symmetry operator
\begin{gather*}Q=\xi^0+x_1\partial_2-x_2\p_1+\ri\eta\end{gather*}
and the following determining equations
 (\ref{5.2}) and (\ref{5.3}):
\begin{gather}\la{ua}\p_\Theta f=af,\quad \p_\Theta V=aV+\dot\eta\end{gather}
where we use the angular variable $\Theta=\arctan(\frac{x_2}{x_1})$.

Like in (\ref{detete}) $a$ and $\dot\eta$ should be constants, and,
by definition, $f$ and $V$ are given by formulae (\ref{sos1}) or (\ref{sos2}).
 We rewrite these formulae using radial and angular variables:
  \begin{gather}\la{sos11}f=F(\tilde r, \Theta)e^{\nu x_3},\
  V=G(\tilde r, \Theta)e^{\nu x_3}\quad \text{ if }\quad Q=P_3+
  \ri\nu t\p_t
\\\la{sos21}f=F(\tilde r, \Theta),\
 V=G(\tilde r, \Theta)+\kappa x_3\quad \text{ if }\quad Q=P_3+\kappa t.\end{gather}

Substituting (\ref{sos11}) into (\ref{ua}) we obtain
\begin{gather*}F=\text{e}^{\sigma\Theta}\tilde F(\tilde r),\quad
G=\text{e}^{\sigma\Theta}\tilde G(\tilde r)\end{gather*} and so
functions $f$ and $V$ are reduced to the form presented in Item 11
of Table 6. Then, substituting (\ref{sos21}) into (\ref{ua}) we
recover functions $f$ and $V$ presented in Item 12 of Table 6.

In analogous way we solve the determining equations corresponding to
the remaining pairs of symmetries represented in (\ref{beba}). As a
result we obtain functions $f$ and $V$ enumerated in Items 13, 14
and 17--21 of Table 6.

The specification of systems admitting  more extended symmetry
algebras presented in Table 5 and Items 22--24 of Table 6 can be
made in analogy with the above with using three dimensional and more
extended subalgebras of algebra so(1,4) which can be found in paper
\cite{Pate}, see also \cite{NZ}. We will not present here the
routine calculations requested to realize this programm.


\begin{thebibliography}{99}
\bibitem{Hag} C. R. Hagen, "Scale and conformal transformations in Galilean-invariant conformal field theory", Phys. Rev. D \textbf{5}, 377--388 (1972).

\bibitem{Nied}
U. Niederer, "The maximal kinematical invariance group of the free
Schr\"odinger equations",
 Helv. Phys. Acta, \textbf{45}, 802--810 (1972).


\bibitem{And}
R. L. Anderson, S. Kumei, C. E. Wulfman,  "Invariants of the
equations of wave mechanics. I.", Rev. Mex. Fis., {\bf 21},  1--33
(1972).


\bibitem{Boy}
C. P. Boyer, "The maximal kinematical invariance group for an
arbitrary potential", Helv. Phys. Acta, {\bf 47}, 450--605 (1974).

\bibitem{11} C. Quesne and V. M. Tkachuk, "Deformed algebras,
position-dependent effective masses and curved spaces: an exactly
solvable Coulomb problem," J. of Phys. A: Math. and Gen. {\bf 37},
4267 (2004).


\bibitem{Cru}Sara Cruz, Y. Cruz and Rosas-Ortiz  Oscar, "Dynamical Equations,
Invariants and Spectrum Generating Algebras of Mechanical Systems
with Position-Dependent Mass," SIGMA  {\bf 9}, 004 (2013).

\bibitem{Mul1} R. Heinonen, E. G. Kalnins, W.
Miller Jr. and E. Subag, "Structure Relations and Darboux
Contractions for 2D 2nd Order Superintegrable Systems",  SIGMA {\bf
11}, 043  (2015).

\bibitem{Mul2} E. G. Kalnins, W. Miller Jr. and E. Subag, "Bocher Contractions of Conformally Superintegrable
Laplace Equations", SIGMA {\bf 12}, 038 (2016).



\bibitem{NZ} A. G. Nikitin and T. M. Zasadko, "Superintegrable systems with
position dependent mass",
     J. Math. Phys. {\bf 56}, 042101 (2015).

    \bibitem{NZ2} A. G. Nikitin and T. M. Zasadko, "Group classification of Schrodinger
     equations with position dependent mass" J. Phys. A: Math.
     Theor. 49 365204  (2016).

     \bibitem{N2} A. G. Nikitin, "Superintegrable and shape invariant
     systems with position dependent mass",
     J. Phys. A: Math. Theor. {\bf 48}  335201   (2015) .

\bibitem{Roz}O. von Roos, "Position-dependent effective masses in
semiconductor theory", Phys. Rev. B {\bf 27}, 7547--7552 (1983).


\bibitem{Pate}
J. Patera and P. Winternitz, "Quantum numbers for particles in de
Sitter space", J. Math. Phys. {\bf 17}, 717--728 (1976).

\bibitem{snob}Libor \^Snobl and Pavel Winternitz, {\it Classification and identification of Lie algebras}, (CRM
Monograph Series, v. 33, 2010)

\bibitem{mur} G. M. Mubarakzianov, "Classification of real structures of
Lie algebras of fifth order", Izvestia Vysshykh Uchebnykh Zavedenii.
Matematika  {\bf 3},  99-106 (1963)

\bibitem{bas}P. Basarab-Horwath, L. Lahno and R. Zhdanov, "The structure of
the Lie algebras and the classification problem  of partial
differential equations", Acta Applicandae Methematica {\bf 69},
43-94 (2001).

\bibitem{boy1}R. O. Popovych,  V. M. Boyko, M. O. Nesterenko and  M. W. Lutfullin,   "Realizations of real low-dimensional Lie algebras", J. Phys. A: Math. Gen. {\bf 36}, 7337-7360 (2003)

\bibitem{boy2} V. Boyko, J. Patera and R. Popovych, R., "Computation of invariants of Lie algebras by means of moving frames", Phys. A: Math. Gen. {\bf 39},  5749-5762 (2006)

\bibitem{ben} D. Ben Daniel and C. Duke, "Space-charge effects on electron
tunneling",  { Phys. Rev} {\bf 152}, 683--92 ( 1966).


 \bibitem{Zhu} Q. Zhu and  H.  Kroemer, "Interface connection rules for
 effective-mass wave functions at an abrupt heterojunction between two
 different semiconductors",  { Phys. Rev. B} {\bf 27}, 3519--27
 (1983).


\bibitem{Cav}  F. Cavalcante, R. Costa Filho, J. Ribeiro Filho,  C.
De Almeida and V. Freire,  "Form of the quantum kinetic-energy
operator with spatially varying effective mass", { Phys. Rev. B}{\bf
55} 1326--28 (1997).

\bibitem{yung} K C. Yung  and J. H. Yee,  "Derivation of the modified Schr\"odinger equation for a
particle with a spatially varying mass through path integrals", {
Phys. Rev. A} {\bf 50} 104–-6 (1994).

\bibitem{mora} R. A. Morrow  and  K. R. Brownstein, "Model Effective
Mass Hamiltonians for Abrupt Heterojunctions and the Associated
Wave-Function-Matching Conditions", {Phys. Rev. B} {\bf 30} 678–-680
(1984).


\bibitem{gor}  T. Gora and  F. Williams, "Theory of Electronic States and
Transport in Graded Mixed Semiconductors", {Phys. Rev.}  {\bf 177}
1179--82 (1969).

\bibitem{li}T. Li  and K. Kuhn,
"Band-offset ratio dependence on the effective-mass Hamiltonian
based on a modified profile of the
${\text{GaAs-Al}_x\text{Ga}_{1-x}\text{As}}$ quantum well",  { Phys.
Rev. B} {\bf 47} 12760--70 (1993).

\bibitem{mustafa} O. Mustafa and S. H. Mazharimousavi,   "Ordering
Ambiguity Revisited via Position Dependent Mass Pseudo-Momentum
Operators",  { Int. J. of Theor. Phys.} {\bf 46} 1786--96 (2007).

\bibitem{khare}
F. Cooper, A. Khare  and U. Sukhatme, "Supersymmetry and Quantum
Mechanics", { Physics Reports} {\bf 251}  267-385 (1995).


\bibitem{LL}J.-M. Lev\'y-Leblond, "Position-dependent effective mass and
Galilean invariance", Phys. Rev. A  {\bf 52}, 1845-1849 (1995).



\bibitem{Pop} A. G. Nikitin and R. O. Popovych, "Group classification of nonlinear Schr\"odinger equations",
Ukr. Math. J. {\bf 53},  1255-1265 (2001).


\bibitem{FN}W. I. Fushchich  and  A. G. Nikitin,  "Higher symmetries and exact
     solutions of linear and nonlinear Schr\"odinger equation",
     {J.  Math. Phys.} {\bf 38}, 5944--59 (1997).

\bibitem{olver}P. Olver, {\it Application of Lie Groups to Differential
Equations} (Springer-Verlag, New York, 2000), 2nd ed.,
electronic version: PJ Olver-2000-books.google.com.

\bibitem{N6}  A.G. Nikitin and R. J. Wiltshire, "Systems of Reaction
     Diffusion Equations and their symmetry properties",
     J.  Math. Phys. {\bf 42},
      1667--88 (2001).

\bibitem{N1} A. G. Nikitin,  "Group classification of systems of non-linear
     reaction-diffusion equations with general diffusion matrix. I.
     Generalized Ginsburg-Landau equations", { J. Math. Analysis and Applications
     (JMAA)} {\bf 324},  615--28 (2006).

\bibitem{Renat}R. Z. Zhdanov and V. I. Lagno, "Conditional symmetry of a
porous medium equation",
 Physica D {\bf 122},  178–-86 (1998).

 \bibitem{gan} L. Gangon and P. Winternitz "Symmetry classes of variable coefficient nonlinear
Schr\"odinger equations", J. Phys. A {\bf 26}  7061–-76 (1993).

\bibitem{popa}A. Bihlo, Elsa Dos Santos Cardoso–Bihlo and R. O. Popovych, "
Complete group classification of a class of nonlinear wave
equations", J. Math. Phys. {\bf 53}, 123515 (2012).



\bibitem{Co}Ian L. Cooper, "An integrated approach to ladder and shift
   operators for the Morse oscillator, radial Coulomb and radial oscillator
   potentials", { J. Phys. A: Math. Gen.} {\bf  26},  1601--1623
   (1993).

   \bibitem{AN1}  J. Beckers, N. Debergh, and A. G. Nikitin, "Reducibility of
     supersymmetric quantum mechanics", Int. J.
     Theor. Phys. {\bf 36}, 1991-2003  (1997).

   \bibitem{AN2}  Niederle J. and  Nikitin A. G., "Extended supersymmetries for
     the Schr\"odinger-Pauli equation",  J. Math.
     Phys, {\bf 40}, 1280-1293 (1999).

   \bibitem{vin1} Genest V. X., Lemay J. M. and Vinet L. "The Hahn
   superalgebra and supersymmetric Dunkl oscillator models", J. Phys.
   A: Math.Theor.
   {\bf 46}, 505204 (2013)

   \bibitem{vin2}H. De Bie, V. X. Genest, J. M. Lemay  and L. Vinet,
   "A
   superintegrable model with reflections on $S^{n-1}$ and higher rank Bannai-Ito
   algebra", ArXiv 1612.07815 (2016).

   \end{thebibliography}
\end{document}